\documentclass[12pt,preprint]{aastex}
\newcommand{\iso}[2]{\hbox{${}^{#1}{\rm #2}$}}
\newcommand{\Msun}{\ensuremath{{M}_{\sun}}}

\shorttitle{AGB stars and Type I PNe}
\shortauthors{Karakas et al.}

\begin{document}


\title{Nucleosynthesis Predictions for Intermediate-Mass AGB Stars: 
Comparison to Observations of Type I Planetary Nebulae\altaffilmark{1}}


\author{Amanda I. Karakas\altaffilmark{2,3}}
\affil{Research School of Astronomy \& Astrophysics, Mt Stromlo Observatory,
Weston Creek ACT 2611, Australia}
\email{akarakas@mso.anu.edu.au}

\author{Mark A. van Raai, Maria Lugaro\altaffilmark{4}}
\affil{Sterrenkundig Instituut, University of Utrecht, Postbus 80000, 3508 TA Utrecht, The Netherlands}
\email{M.A.vanRaai@students.uu.nl, m.a.lugaro@uu.nl}

\author{N. C. Sterling}
\affil{NASA Postdoctoral Program Fellow, Goddard Space Flight Center, Code 662, Greenbelt, MD 20771}
\email{Nicholas.C.Sterling@nasa.gov}

\and

\author{Harriet L. Dinerstein}
\affil{University of Texas, Department of Astronomy, 1 University Station, C1400, Austin, TX 78712-0259}
\email{harriet@astro.as.utexas.edu}


\altaffiltext{1}{This paper includes data taken at The McDonald Observatory of The University of Texas at Austin}
\altaffiltext{2}{Physics Division, Argonne National Laboratory, Argonne, IL 60439-4843}
\altaffiltext{3}{Department of Astronomy and Astrophysics, 
University of Chicago, 5640 S. Ellis Avenue, Chicago, Illinois 60637}
\altaffiltext{4}{Centre for Stellar \& Planetary Astrophysics, Monash University, Clayton VIC 3800, Australia}


\begin{abstract}
Type I planetary nebulae (PNe) have high He/H and N/O ratios
and are thought to be descendants of stars with initial masses 
of $\sim 3$ -- 8$\Msun$. These characteristics indicate that 
the progenitor stars experienced proton-capture nucleosynthesis 
at the base of the convective envelope, in addition to
the $slow$ neutron capture process operating in the He-shell
(the $s$-process).
We compare the predicted abundances of elements up to Sr from
models of intermediate-mass asymptotic giant branch 
(AGB) stars to measured abundances in Type I PNe. In particular, 
we compare predictions and observations for the light
trans-iron elements Se and Kr, in order to constrain convective mixing
and the $s$-process in these stars.  A partial mixing zone is 
included in selected models to explore the effect of a \iso{13}C 
pocket on the $s$-process yields.  The solar-metallicity 
models produce enrichments of [(Se, Kr)/Fe] $\lesssim~0.6$, 
consistent with Galactic Type I PNe where the observed enhancements
are typically $\lesssim 0.3$ dex, while lower metallicity models
predict larger enrichments of C, N, Se, and Kr.
O destruction occurs in the most massive models but it is 
not efficient enough to account for the $\gtrsim 0.3$~dex 
O depletions observed in some Type~I PNe.
It is not possible to reach firm conclusions regarding the neutron 
source operating in massive AGB stars from Se and Kr abundances 
in Type I PNe; abundances for more $s$-process elements 
may help to distinguish between the two neutron sources.
We predict that only the most massive ($M \gtrsim 5\Msun$) 
models would evolve into Type I PNe, indicating that extra-mixing
processes are active in lower-mass stars (3--$4\Msun$), if
these stars are to evolve into Type~I PNe.
\end{abstract}




\section{Introduction}

After the thermally-pulsing AGB (TP-AGB) phase is terminated, low to 
intermediate mass stars ($\sim 0.8$ to 8$\Msun$) evolve to become 
post-AGB stars and possibly planetary nebulae (PNe), if the ejected
envelopes have sufficient time to become ionized by the hot central 
stars before dissipating into the interstellar medium. For recent 
reviews of TP-AGB and post-AGB stars see \citet{herwig05} and
\citet{vanwinckel03}, respectively.  
The illuminated PN is comprised of material
from the deep convective envelope, hence nebular abundances should 
reveal information about the efficiency of mixing events and chemical 
processing that took place during previous evolutionary phases,
in addition to the initial composition of the parent star
\citep{dopita97,karakas03a}.

Briefly, during the TP-AGB phase the He-burning shell becomes 
thermally unstable every $10^{4}$ years or so,
depending on the core mass. The energy from the thermal pulse 
(TP) drives a convective zone in the He-rich intershell, 
that mixes the products of He-nucleosynthesis within this region. 
The energy provided by the TP expands the whole star, pushing 
the H-shell out to cooler regions where it is almost 
extinguished, and subsequently allowing the convective 
envelope to move inwards (in mass) to regions previously 
mixed by the flash-driven convective zone. 
This inward movement of the convective envelope is known as 
the third dredge-up (TDU), and is responsible for enriching
the surface in \iso{12}C and other products of He-burning,
as well as heavy elements produced by the $s$ process. 
Following the TDU, the star contracts and the H-shell is 
re-ignited, providing most of the surface luminosity for 
the next interpulse period.
In intermediate-mass AGB stars with initial masses 
$\gtrsim 4\Msun$, the base of the convective 
envelope can dip into the top of the H-burning shell, 
causing proton-capture nucleosynthesis to occur there 
\citep[hot bottom burning, HBB; see][for a review]{lattanzio96}.

Abundances can be obtained from PN spectra for a number of 
elements including C, N, O, S, Cl and the noble gases He, Ne, and
Ar \citep{aller83,kingsburgh94,dopita97,stanghellini00,leisy06}; 
and heavy elements synthesized by the $s$ process, including 
Ge, Se, Kr, Xe, Ba, and Br 
\citep[][hereafter SD08]{pequignot94, dinerstein01a, 
sterling02, sharpee07, sterling07, sterling08}.
Whereas the metallicity of a star is usually taken from the
abundance of iron, for ionized nebulae oxygen is generally the
primary index of metallicity. However, O is not an ideal 
metallicity index for PNe, because it can be destroyed by CNO 
cycling and/or synthesized during helium burning. 
Consequently, Ar is sometimes used in place of O \citep[][SD08]{leisy06},
under the assumption that it remains unchanged by AGB nucleosynthesis.
Zinc is a better indicator of iron-group element abundances 
in nebulae than Fe itself, because Zn does not condense into 
dust as easily as Fe \citep{york82,sembach95,welty99} and can 
be observed in PNe \citep{dinerstein01,dinerstein04}.  
However, Zn is at the beginning of the $s$-process chain and 
might be produced to some extent in AGB stars.
In order to interpret the elemental abundances in PNe it is 
important to determine which elements are produced, destroyed 
or left unaltered by AGB nucleosynthesis.

The set of elements heavier than Fe that can be detected 
in PNe include all even-numbered elements from $Z$ = 30 -- 36, 
which lie at the light-element end 
of the $s$-process distribution, just
below the first $s$-process peak at Sr, Y, and Zr ($Z$ = 38 -- 40).
In \citet{karakas07a} we attempted to use Ge abundances
derived from a small sample of PNe \citep{sterling02} to
try and constrain the amount of TDU mixing after the final
few TPs. There a hint was found that efficient TDU at the tip 
of the AGB may be required in order to match the 
abundances; however, stronger conclusions were precluded
by model uncertainties, the small PNe sample size, 
and the large uncertainties of the Ge abundance determinations.
In the present study we proceed to analyze Se and Kr 
abundances from a much larger 
sample of 120 Galactic PNe from SD08, where the
$s$-process enrichments were investigated for correlations 
with PN morphology and other nebular and stellar 
characteristics, as outlined in detail in \S\ref{obs}.
We also use PNe abundances of C, N, and O along with 
the abundances of post-AGB stars for comparison to the
stellar models.

The composition of Type I PNe provides important constraints 
on nucleosynthesis and mixing in intermediate-mass AGB stars.
These PNe tend to exhibit bipolar morphologies and have high He/H
and N/O ratios characteristic of proton-capture 
nucleosynthesis \citep{stanghellini06}, suggesting 
they are descendants of AGB stars with initial masses 
$\sim$3 -- 8 $\Msun$ 
\citep{peimbert78,kingsburgh94,stanghellini06}. Low-mass 
AGB stars, on the other hand, are defined as having 
initial masses $\sim$1 -- 3$\Msun$ and tend to show carbon 
and $s$-process element enrichments \citep{busso01,travaglio04}.
In the Milky Way, the spatial distribution and kinematics of 
Type I PNe indicate that they are a young population 
\citep[e.g.,][]{corradi95}. This supports their identification 
with intermediate-mass stars, which have relatively short 
evolutionary lifetimes. 
In the Galaxy, intermediate-mass AGB stars are difficult 
to identify owing to a lack of reliable distances. 
This has resulted in a paucity of observational evidence for 
constraining stellar models, and especially the efficiency 
of the TDU in intermediate-mass AGB stars, which is still not 
well determined 
\citep{karakas02,ventura02,herwig04a,stancliffe04b}.
\citet{garcia06} identified several Galactic intermediate-mass 
AGB stars within a sample of OH/IR stars (i.e., bright O-rich 
giants with large infrared excesses). The large enhancements
of Rb found by these authors, combined with the fact that 
these stars are O-rich, support the prediction that HBB 
and an efficient TDU has occurred in these stars.
At lower metallicities, constraints are provided by observations 
of luminous O-rich AGB stars in the Large and Small Magellanic 
Clouds (LMC and SMC, respectively) that are rich in Li and 
$s$-process elements \citep{wood83,smith89,plez93}.
 
We begin in \S\ref{obs} with a summary of observational results 
for $n$-capture elements in PNe. We review the numerical method 
and present the stellar models in \S\ref{models}, and outline 
the results in \S\ref{results}. In \S\ref{discussion}, 
we discuss our findings and their implications, and
in \S\ref{conclusions} we summarize our conclusions.

\section{Observational Constraints} \label{obs}

The low cosmic abundances of \emph{n}-capture elements cause their
spectroscopic features to be very weak.  For that reason, trans-iron
elements were not identified in the spectrum of a PN until
1994 \citep{pequignot94}.
However, in the last few years there has been considerable progress
in the detection of emission lines from and derivation of
abundances for trans-iron elements in PNe
\citep[e.g.,][SD08]{dinerstein01,sterling07,sharpee07}.

The most comprehensive study is that of SD08, who
conducted the first large-scale survey of \emph{n}-capture elements
in PNe. They determined Kr and Se abundances in 120 PNe from the
near-infrared emission lines [\ion{Kr}{3}]~2.199~$\mu$m and
[\ion{Se}{4}]~2.287~$\mu$m identified by \citet{dinerstein01a}, 
and corrected for the abundances of unobserved Se and Kr ions using
analytical formulae derived from a grid of photo-ionization models
\citep{sterling07}. This survey increased the number of PNe with 
determined \emph{n}-capture element abundances by nearly an order of 
magnitude, enabling a detailed study of
\emph{s}-process enrichments in PNe as a population,
as well as an assessment of enrichment patterns in different 
classes of PNe. In order to normalize the abundances of the
\emph{s}-process products to the initial metallicity of 
each PN, the Se and Kr abundances were compared to those of O 
and Ar reported in the literature. Oxygen was utilized as a 
reference element for PNe with low-mass progenitors, since its 
abundance is generally the most reliably determined in PNe,
while Ar was used for objects with
higher-mass central stars that may have experienced O destruction
during HBB.  Overall, 41 of the 94 PNe with derived
Se and/or Kr abundances or meaningful upper limits
were found to be enriched in
these elements, with the average [Se/(O,~Ar)]$ = 0.31$ and
[Kr/(O,~Ar)]$= 0.98$\footnote{We use the notation [Se/(O,~Ar)] 
and [Kr/(O,~Ar)] of SD08 to emphasize the use of different reference
elements for PNe with different progenitor masses, where
$\mathrm{[X/Y]}=\mathrm{log_{10}(X/Y)}-\mathrm{log_{10}(X/Y)}_{\odot}$.}.
SD08 interpreted these enrichments as evidence for \textit{in situ}
\emph{s}-process nucleosynthesis and dredge-up in PN progenitor
stars.

\begin{figure}
\epsscale{0.85}
\plotone{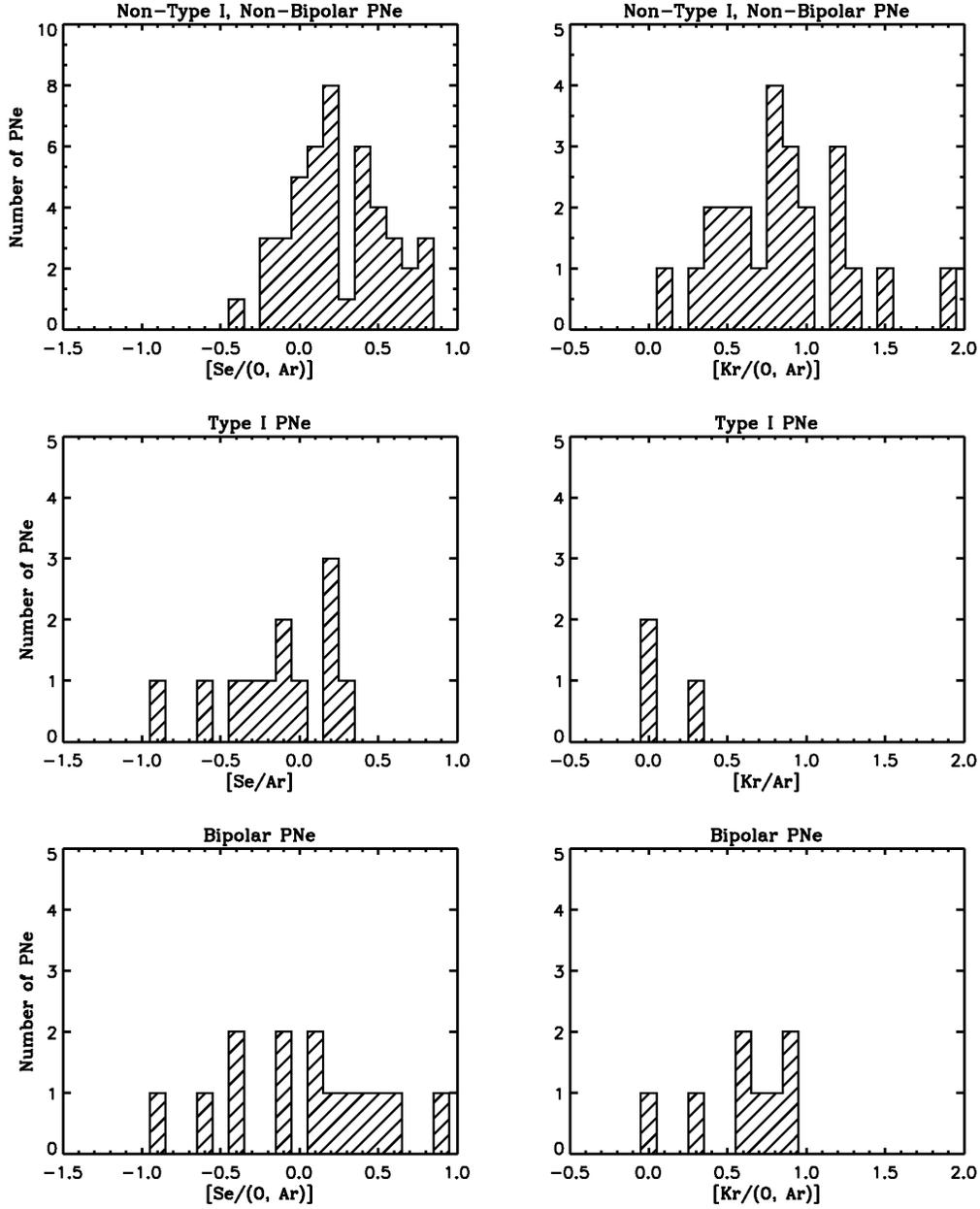}
\caption{Histograms of Se and Kr abundances from SD08, separated into 
0.1~dex bins, for non-bipolar, non-Type~I PNe (top); Type~I PNe (middle); 
and bipolar PNe (bottom).  Data are shown only for objects in which 
Se and/or Kr emission was detected.  Type~I and (to a lesser extent) 
bipolar PNe exhibit smaller enrichments than PNe with less massive
progenitor stars. \label{sekrhistogram}}
\end{figure}

SD08 observed significant numbers of PNe with
intermediate-mass progenitor stars, as evidenced
by their Type I compositions
and/or bipolar morphologies (29 and 28 objects
respectively; see their Table 15).
They found that Type~I and bipolar PNe
exhibit smaller Se and Kr enrichments
on average than other PNe.  In Figure~\ref{sekrhistogram} 
we display [Se/(O,~Ar)] and [Kr/(O,~Ar)], separated into 
0.1~dex bins, for non-Type~I, Type~I, and bipolar PNe.
The distribution of Se and Kr enrichments is clearly
skewed to smaller values for Type~I PNe than other objects,
and to a lesser extent the same is true of bipolar PNe.  
Notably, all of the Type~I PNe observed by SD08 exhibit modest
(less than a factor of two, or 0.3~dex)
or no Se and Kr enrichments. These values are
significantly lower than the average values for
non-Type~I PNe: 0.36~dex for [Se/(O,~Ar)], and 1.02~dex for
[Kr/(O,~Ar)]. 
In contrast, while some bipolar PNe are among the objects
with the lowest [Se/(O,~Ar)] and [Kr/(O,~Ar)], others 
display enrichments approaching a factor of 10.

\begin{figure}
\epsscale{0.8}
\plotone{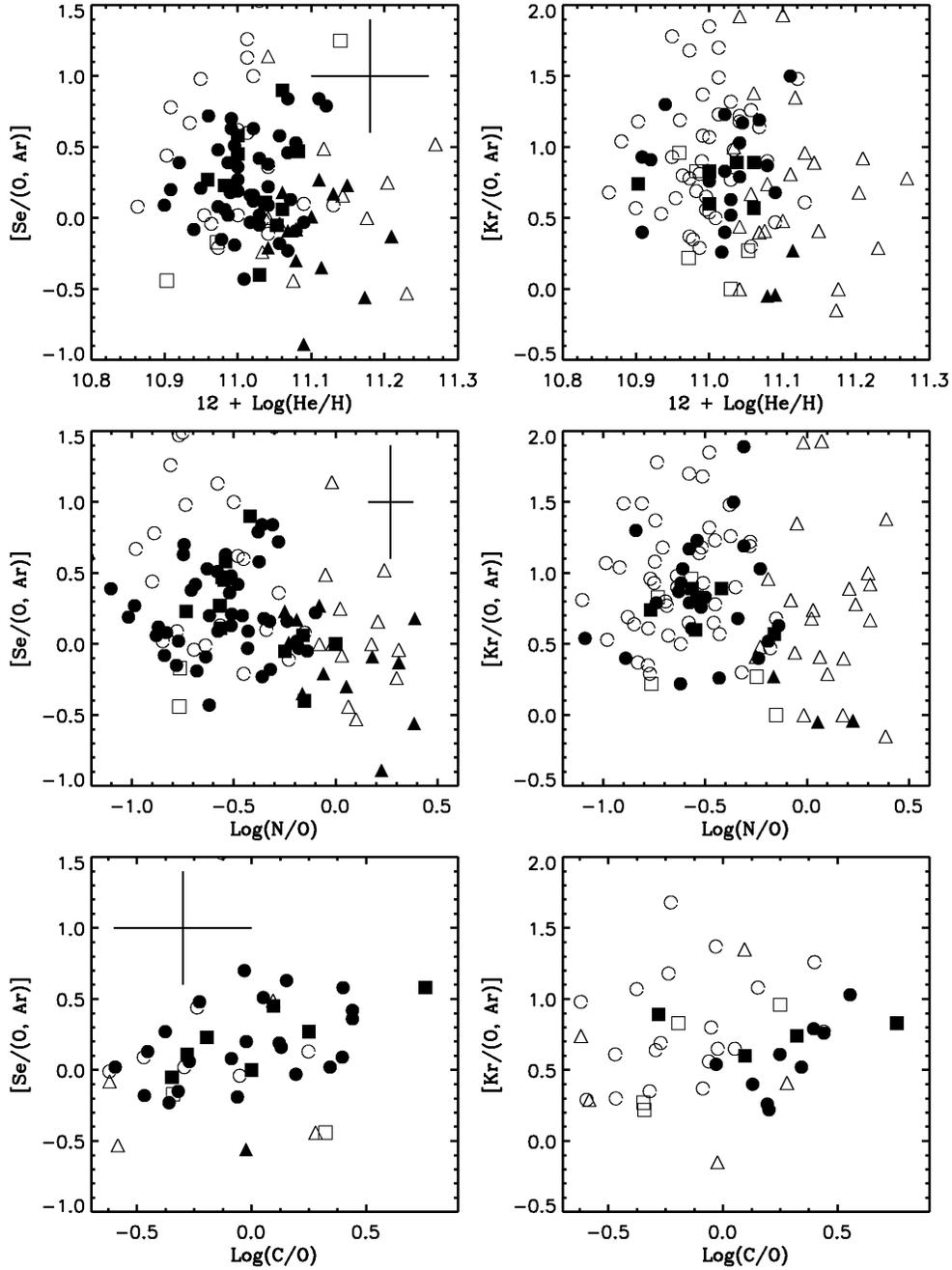}
\figcaption{The Se and Kr abundances of PNe plotted against He/H (top panels), 
N/O (middle panels), and C/O (bottom panels).  The triangles correspond to 
Type~I PNe, the boxes are non-Type~I bipolar PNe, and circles represent all 
other PNe from the sample of SD08.  Se and Kr upper limits are depicted as 
open symbols.  Representative error bars to the abundances are shown in the 
left hand panels.  The observed correlations show that Type~I PNe generally display 
smaller \emph{s}-process and C enrichments than other PNe.  While many 
bipolar PNe follow the same trends, some are significantly enriched in 
C, Se, and Kr, suggesting that they arise from less massive stars.
\label{sekrcorrelations}}
\end{figure}

Figure~\ref{sekrcorrelations} shows [Se/(O,~Ar)] and 
[Kr/(O,~Ar)] as a function of the logarithmic He/H, N/O, 
and C/O abundances of PNe from the SD08 sample, including 
upper limits (open symbols) for objects
without detected Se and/or Kr emission.  This figure
not only illustrates the lower average Se and Kr abundances of
Type~I PNe relative to other objects, but also that upper limits 
to the Se and Kr abundances in many Type~I PNe allow for 
at most marginal \emph{s}-process enrichments.
SD08 found that \emph{s}-process enrichments are
positively correlated with C/O in non-Type I PNe 
(bottom panels of Figure~\ref{sekrcorrelations}),
a result that agrees with studies of AGB \citep{smith90a,abia02} 
and post-AGB stars \citep{vanwinckel03}. Type I PNe 
often exhibit low C/O ratios ($< 1$), in agreement with 
predictions from intermediate-mass stars with HBB. 
HBB converts dredged-up $^{12}$C into $^{14}$N, thereby
preventing or delaying the formation of C-rich AGB stars
\citep{frost98a}. 
On the other hand, some bipolar PNe display significant 
enrichments of both C and \emph{s}-process nuclei, as well as
non-Type~I compositions.  These objects may be
descendants of low-mass binary star systems in which the bipolar
morphology is a result of binary interactions
\citep{soker97,balick02}.
This suggests that some bipolar PNe are not descendants of
intermediate-mass stars, and that morphology is not as reliable an
indicator of progenitor mass as is chemical composition.

\citet{sharpee07} derived \emph{n}-capture element abundances in
five PNe from high resolution optical spectra.  They identified
emission lines of several \emph{n}-capture elements, including Br,
Kr, Xe, Rb, Ba, and Pb in each of these objects, and derived
abundances for Br, Kr, and Xe.  They found that Kr and Xe are
enriched in three of the five observed PNe, by 0.3 to 0.9~dex.
The two objects in their sample which do not exhibit \emph{s}-process
enrichments, NGC~2440 and IC~2501, are both Type~I PNe
according to the classification scheme of \citet{peimbert78}.
The [Kr/Ar] and [Xe/Ar] values of these two objects are consistent
within the uncertainties with the solar values, except for Xe in
NGC~2440, which may be mildly subsolar.  These results for [Kr/Ar]
are consistent with the findings of SD08 for their 
larger sample of Type~I PNe.

The near-infrared [\ion{Kr}{3}] and [\ion{Se}{4}] lines 
utilized by SD08 have also been detected in two PNe belonging 
to metal-poor Milky Way dwarf satellite galaxies
\citep[][Dinerstein et al., in preparation]{wood06}.
Such observations can provide important constraints on 
AGB models and nucleosynthetic yields for subsolar 
metallicities. One of the observed objects, LMC~SMP~62,
is N-rich (N/O~$\sim$~0.5) and hence considered a
Type~I PN in the context of the LMC, where the
threshold value of N/O for Type~I objects is
taken to be lower than for Galactic PNe
\citep{dopita92,leisy06}.
While the [\ion{Kr}{3}] line is not detected 
(and does not provide a useful abundance 
limit), the [\ion{Se}{4}] line yields a Se abundance 
$\sim$~15\% solar (Dinerstein et al., in preparation).
Not only is Se not enriched, it is actually \emph{deficient} 
relative to the $\alpha$-elements; the abundances of O, Ne, S, 
and Ar in SMP 62 are $\sim$~30--40\% solar
%%
%% begin footnote
\citep{leisy06,bernard08}\footnote{Using $\it{Spitzer}$ 
spectra, \citet{bernard08}
find higher Ne/H than \citet{leisy06}.
However, they adopt a higher value for the solar
Ne abundance, and consequently their ratio to solar is similar.
They also derive a low value for S/H, but the significance
of this result is unclear, since these authors also find
that Galactic PNe have subsolar sulfur abundances
\citep{pottasch06}.}.
%% End footnote
%%
This is reminiscent of the abundance pattern in
LMC red giants, where light $s$-process elements 
such as Y and Zr are markedly deficient relative to iron
and some $\alpha$ elements \citep{pompeia07,muccia08}.
In contrast, Hen~2-436, a PN in the Sagittarius dSph galaxy,
displays Se and Kr abundances as high as those
for some of the most enriched Galactic objects shown in
Figure~\ref{sekrhistogram}. However, Hen~2-436 is C-rich
rather than N-rich, and therefore is a non-Type~I PN,
with an estimated initial stellar mass of
$\sim$~1.2~$\Msun$ \citep{dudziak00,zijlstra06}.

Zinc is detectable in PNe via the [\ion{Zn}{4}] 
3.625 $\micron$ fine-structure line identified by 
\citet{dinerstein01}. This line has recently been observed
in about a dozen PNe belonging to the Milky Way
and its dwarf satellites 
\citep[Dinerstein, Geballe, \& Sterling, 
in preparation, hereafter DGS08]{dinerstein07}.
Two of these objects exhibit Type~I compositions:
LMC~SMP~62, discussed above; and the Galactic PN 
M~1-40 \citep{gorny04}.  The gaseous Zn abundance 
in M~1-40 is $\sim$~80\% solar, which is
consistent with its approximately solar Ne and S abundances.
Even allowing for modest depletion into dust,
which would raise the total Zn abundance by 0.1--0.2~dex
(assuming conditions typical of the warm ISM; see \S 1),
it does not appear that Zn is significantly enriched in M~1-40.
In LMC~SMP~62, the measured Zn abundance is
lower relative to solar than the abundances of $\alpha$ elements
such as O, Ne, and S \citep{dinerstein07} by a factor
very similar to the deficiency in Se.
In this context, it is interesting
that several of the iron-group elements,
including Ni and Cu (which are adjacent to Zn),
display behavior similar to light $n$-capture elements, 
which are also deficient relative to Fe in LMC giants 
\citep{pompeia07,muccia08}.

\section{The stellar models} \label{models} 

The numerical method we use has previously been described in detail
\citep{lugaro04,karakas06a}. 
We first compute the stellar structure using 
the Mt Stromlo Stellar Structure code \citep{lattanzio86}, where
each model is evolved from the zero-age main sequence to near the
tip of the TP-AGB. For the models considered in this study, we do not
evolve to the post-AGB phase due to convergence difficulties 
that occur when the envelope mass is reduced below $\sim 1.5\Msun$;
see \citet{karakas07b} for more details.  The structure is
then used as input into a post-processing nucleosynthesis code
where we obtain abundances for many more species (up to 166) 
than are included in the structure model (6 species).

The stellar models we computed (see Table~\ref{stellarmodels}) cover a 
mass range of 3.0--$6.5 M_{\odot}$ and metallicities 
0.2--1.0~$Z_{\odot}$.  The metallicities were chosen to reflect the 
composition of most of the PNe in the SD08 sample, where O and Ar 
abundances were used as metallicity indicators.  
We include a 2.5$\Msun$, $Z=0.008$ model as an example 
of a low-mass, low-metallicity progenitor
that would lead to a non-Type I PN composition.
All models use scaled solar abundances from 
\citet{anders89} except for the $Z=0.012$
models, which are computed with the revised solar 
elemental abundances from \citet*{asplund05} for comparison.
\citet{reimers75} mass loss was used on the first giant branch, 
with the parameter $\eta=0.4$, and the \citet{vw93} mass loss 
on the AGB. For comparison, we include results for a 5$\Msun$, 
$Z=0.02$ model computed with Reimers mass loss on the AGB, 
with the parameter $\eta = 3.5$. 

In Table~\ref{stellarmodels} we summarize structural information 
about the models. Most of the structure calculations are
taken from \citet{karakas07b}, with
the $Z=0.012$ models discussed previously in \citet{karakas07a}. 
In Table~\ref{stellarmodels} we present the initial mass and metallicity, $Z$;
the total number of TPs computed; the maximum temperature in the He-shell,
$T_{\rm He}^{\rm max}$; the maximum temperature at the base of the 
convective envelope, $T_{\rm bce}^{\rm max}$; the total mass dredged into
the envelope during the TP-AGB, $M_{\rm dred}$; the final envelope 
mass $M_{\rm env}$; and whether or not HBB occurs.
In Table~\ref{stellarmodels}, masses are given in solar units,
temperatures in millions of kelvins, metallicity $Z$ is the fraction
by mass of metals, and elemental and isotopic ratios are in
terms of number ratios.

\subsection{Type I planetary nebulae and HBB}

Observations suggest that the Type I PN progenitor mass 
range is $\sim$~3 to 8~$\Msun$ \citep[e.g., ][]{peimbert90,stanghellini06}.
For the purpose of this study we consider models with 
masses between 3 to 6.5~$\Msun$, 
noting that more massive AGB stars would likely evolve too 
quickly to form observable PNe.  This mass range encompasses the
full range of AGB nucleosynthetic behavior: (1) models that
experience the TDU but not HBB, and (2) models that undergo the 
second dredge-up (SDU) and HBB in addition to the TDU.
From inspection of Table~\ref{stellarmodels}, the 2.5, 3, and 
4~$\Msun$ models fall into the first category while the 5, 6, 
and 6.5~$\Msun$ models fall into the latter category. 
The more massive objects that experience both the SDU
and HBB are often defined as intermediate-mass AGB
stars \citep[see][]{karakas06a,herwig05}.

Only models with HBB (i.e., models with initial masses 
$m \gtrsim 5 \Msun$) would evolve into 
Type I PNe.  Hence efficient extra mixing between 
the base of the convective 
envelope and a region hot enough to allow some CN 
cycling is presumably operating to produce the 
observed Type I nucleosynthethic signature of high He/H 
and N/O ratios in the lower-mass $\sim$ 3--4$\Msun$ stars.
The physical mechanism that could cause such efficient 
extra mixing is not known, although stellar
rotation is one possible candidate (see below). 
It is probably not thermohaline mixing which is 
only efficient in stars less massive than $2\Msun$ 
\citep{eggleton06,charbonnel07}.
The idea that extra-mixing processes are operating 
in low-mass carbon-rich AGB stars is supported by 
observational evidence, including lower than predicted 
\iso{12}C/\iso{13}C ratios observed in N-type AGB 
stars \citep{abia97}. 
The C, O, and Al isotopic ratios found in meteoritic 
silicon carbide (SiC) and oxide grains from AGB stars 
also point 
toward some sort of extra-mixing process operating 
in the parent stars  \citep{busso99,nollett03,zinner06,zinner08}.
In these cases, the mixing should not be so efficient to 
prevent the formation of a carbon-rich atmosphere.

Some fraction of Type I and bipolar PNe presumably formed 
from low-mass stars
via binary evolution \citep{soker97,balick02,moe06}, 
although it is unclear how this would lead to the high He/H 
and N/O ratios observed.  Two-dimensional simulations of a 
star with a point-mass companion in a very close orbit 
suggested that close binary evolution
has little effect on the deep interior structure,
despite producing large distortions to the outer layers \citep{deupree05}.
However, these simulations did not go beyond the point 
where the stars begin to fill their Roche lobes. 
Rapid stellar rotation in combination with magnetic fields 
has also been proposed as a mechanism to shape bipolar PNe
and to produce the overabundances of He and N in single
intermediate-mass progenitors ($M \gtrsim 3\Msun$)
\citep{calvet83,gorny97,garcia-segura99,dobrincic08}.
Further modeling is required to address how effectively 
binary interactions, rotation, and/or magnetic fields 
can affect the internal structure and nucleosynthesis of 
AGB stars.

\begin{figure}
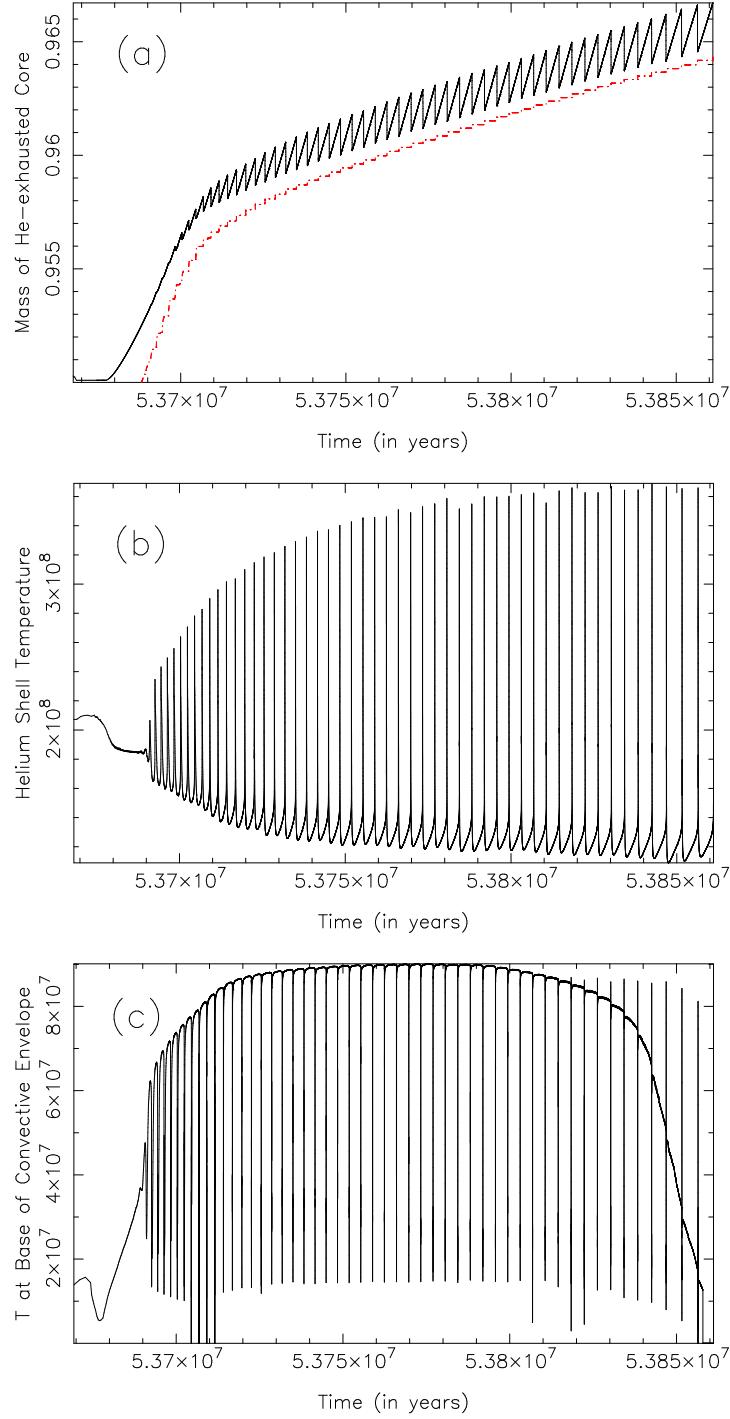

\begin{center}
\begin{tabular}{c}
\includegraphics[width=6cm,angle=270]{f3a.ps} \\
\includegraphics[width=6cm,angle=270]{f3b.ps} \\
\includegraphics[width=6cm,angle=270]{f3c.ps}
\end{tabular}
\caption{The temporal evolution of the (a) hydrogen (solid line) and 
helium-exhausted (dashed line) cores, (b) the He-shell 
temperature and (c) the temperature at the base of the convective 
envelope for the 6.5$\Msun$, $Z = 0.012$ model during the TP-AGB.
\label{heshelltemp}}
\end{center}
\end{figure}

In Fig.~\ref{heshelltemp} we plot the temporal evolution 
of the core, and the temperature in the He-shell 
and at bottom of the convective envelope for the 6.5$\Msun$, 
$Z=0.012$ model.   The mass of the H-exhausted core shows
evidence of efficient TDU whereas the He-shell temperature
shows that the majority of the TPs reach peak temperatures
greater than 300 $\times 10^{6}$K, the temperature at which
the \iso{22}Ne neutron source is activated. 
HBB prevented several models in Table~\ref{stellarmodels}
from becoming carbon rich, and also resulted in 
low \iso{12}C/\iso{13}C ratios that approach the 
equilibrium ratio of $\sim 4$.
The He/H and N/O ratios are also higher for models with
HBB compared to the lower mass models.  
HBB is more efficient at lower metallicities for a given mass
owing to the higher temperatures reached at the base of
the convective envelope. 
For example, the 6.5 M$_{\odot}$, $Z=0.012$ model 
(corresponding to the Asplund et al. solar 
abundances) displays a higher N/O ratio than the 
$Z=0.02$ model of the same mass. The final N abundance is 
similar in each of these two models, despite a 
difference of a factor of 1.86 in the initial N abundances, 
showing that CNO processing is more efficient in the 
low-metallicity case.
%% 
%% initial N abundance  0.012 4.2364296E-05  1.5575006E-07 
%%                      0.02  7.8990866E-05  2.9115438E-07
%% initial O            0.012 3.2184130E-04
%%                      0.02  6.0041249E-04
%% initial C            0.012 1.7133198E-04
%%                      0.02  2.5297439E-04

\subsection{The nuclear network}
 
The most important addition to our models for this study 
is the extension of the nuclear network to include elements 
heavier than iron up to niobium (M. van Raai et al., in preparation).
We have also made an update of our reaction-rate library -- 
starting from the library described in \citet{karakas07a}
we have included neutron capture cross sections from the 
\citet{bao00} compilation.
We include 166 species from protons to sulfur and 
iron through to Nb, with a total of 1285 reaction rates 
corresponding to all $\beta$-decay, $p$, $\alpha$, and $n$-capture 
reactions on all species in the network. We also have two other 
networks that we use to obtain information for the S and Ar isotopes.
The first has 125 species and includes all stable species from H 
to \iso{62}Ni, while the second has 156 species, and includes all
stable species from H to arsenic \citep{karakas07a}. 

As done in \citet{karakas07a}, we use a 
``double neutron-sink'' 
description \citep*{herwig03} to account for neutron captures on 
species heavier than that at the end of the network 
(\iso{97}Nb in this case).  The two artificial species are 
linked by the reactions \iso{97}Nb($n, \gamma$)\iso{98}g and 
\iso{98}g($n,L$)\iso{98}g, where \iso{98}g has an initial abundance 
equal to the sum of the solar abundances from Mo to Bi.   
The second artificial particle, $L$, is 
equivalent to the number of neutrons captured beyond Nb.
The ratio ($L$/\iso{98}g) is a 
description of the neutrons captured per seed nucleus and could in 
principle be related to the $s$-process distribution.
Note that in the 156 and 166 networks, the species \iso{34}S is 
the sum of all species from \iso{34}S to Mn, and the reaction 
$^{34}$S(n,$\gamma$)$^{35}$S is assigned an averaged cross 
section value in order to represent all nuclei from 
$^{34}$S to Mn.     

\subsection{Neutron sources in intermediate-mass AGB stars}

The \emph{s} process is driven by the production of free 
neutrons that are subsequently captured by Fe-peak seed nuclei 
to form heavier elements.  The \iso{22}Ne($\alpha, n$)\iso{25}Mg 
reaction was first suggested as a neutron source in stars by 
\citet{cameron60}. Later, \citet{truran77} and \citet{cosner80} 
suggested it was the dominant source in intermediate-mass AGB 
stars, because the He-shells of these stars reach 
high enough temperatures ($T \gtrsim 300 \times 10^{6}\,$K) 
to allow for this reaction to be efficiently activated 
(see the middle panel of Fig.~\ref{heshelltemp} as an example). 
These temperatures are reached only in the last few 
TPs of a lower mass star.
\citet{truran77} found that the \iso{22}Ne source results in
enhanced levels of the $n$-rich element Rb owing to 
the high neutron densities (up to $\sim 10^{13}$ 
neutrons/cm$^{3}$), as well as increases in the heavy Mg 
isotopes \iso{25}Mg and \iso{26}Mg 
from the competing reactions \iso{22}Ne($\alpha,n$)\iso{25}Mg 
and \iso{22}Ne($\alpha,\gamma$)\iso{26}Mg 
\citep{kaeppeler94,karakas06a}. \citet{fenner03} compared 
results from a chemical evolution model with observations of 
the neutron-rich Mg isotopes and concluded that an extra 
production site besides Type II supernovae was necessary,
and that massive AGB stars are a good candidate for such a 
site. There is, however, some uncertainty at what Galactic 
epoch AGB stars started contributing to the chemical enrichment 
of the Galaxy \citep{simmerer04,melendez07}. 
It is also unknown to what extent intermediate-mass stars 
contribute to the Galactic inventory of $s$-process 
elements \citep[e.g.][]{travaglio04}. 

The other potential source of neutrons in AGB stars is the 
\iso{13}C($\alpha,n$)\iso{16}O reaction, which operates at 
lower temperatures ($T \gtrsim 90 \times 10^{6}$) than the 
\iso{22}Ne source.   Observational and theoretical evidence 
suggests this is the dominant neutron source in low-mass
AGB stars \citep{smith87,gallino98}.
To operate efficiently, this reaction requires more 
\iso{13}C than is left over from CN cycling in the H-shell; 
hence some mechanism to mix protons from the H-rich envelope 
into the intershell is needed to produce the extra \iso{13}C.
In our models, protons are mixed into the intershell 
region by artificially adding a partial mixing zone (PMZ) 
at the deepest extent of each TDU.  These protons
are quickly captured by the abundant \iso{12}C to form 
\iso{13}C and \iso{14}N, resulting in the formation of 
a \iso{13}C pocket. In the \iso{13}C pocket, neutrons 
are liberated by the reaction
\iso{13}C($\alpha,n$)\iso{16}O during the interpulse 
period \citep{straniero95}, in contrast to the \iso{22}Ne 
source, which operates during TPs.
The timescales for neutron production during the interpulse are 
much longer ($\gtrsim 10^{3}$ years) than during the convective 
pulse ($\sim 10$ years), resulting in much lower neutron 
densities ({\bf $\sim 10^{7}$} neutrons/cm$^{3}$) than the \iso{22}Ne
source.  Together, the timescale for neutron
production and the neutron source determine the resulting
$s$-process element distribution. The details of how the 
\iso{13}C pocket forms and its extent in mass in the 
He-intershell are still unknown, although various mechanisms 
have been proposed, including convective overshoot, rotation, 
and gravity waves; see \citet{herwig05} for a discussion of 
the relative merits of each mechanism.

The importance of the \iso{13}C pocket and the 
\iso{13}C($\alpha,n$)\iso{16}O reaction for nucleosynthesis in 
intermediate-mass stars is neither clear nor well studied. 
Adding a PMZ artificially into massive AGB models
has little effect on the nucleosynthesis of nuclei 
lighter than Fe \citep{karakas06a} and on the 
production of Ge in 5$\Msun$, $Z=0.02$ models \citep{karakas07a}. 
The first point can be understood by 
noting that the \iso{13}C pocket is about $\sim$~10\% 
of the mass of the He-intershell. 
Since the He-intershell is already smaller in 
massive AGB stars by about one order of magnitude compared to 
lower-mass stars, the inclusion of the PMZ has little effect 
on the surface abundances of light elements, which are relatively 
abundant in comparison to heavy elements produced via the $s$ process.  
The effect of the \iso{13}C pocket on the production
of $s$ nuclei is less clear, given their low initial abundances.
Se and Kr elemental abundances derived from Type I PN spectra,
in combination with observations of the elements Rb and Zr in 
massive AGB stars in the Galaxy, LMC, and SMC, may provide a 
way of distinguishing between the relative importance of the 
two neutron sources in intermediate-mass AGB stars.

For the 3, 4, 5, and 6.5~$\Msun$ solar composition models
we present results with and without a PMZ 
(see Table~\ref{abund} for the PMZ masses used in each model). 
No PMZ was included in any of the low-metallicity models, 
with the exception of the 2.5 $M_{\odot}$, $Z=0.008$ model,
where we only include results with a PMZ.
The PMZ masses were chosen such that the resulting 
\iso{13}C pockets were $\sim 10$\% to 15\% of the 
mass of the He-intershell, consistent with the \iso{13}C
pocket masses used by \citet{gallino98}.
The method we use to include a PMZ is described in 
\citet{lugaro04}, and is similar to that used
by \citet{goriely00}. At the deepest extent of each TDU episode, 
we add an exponentially decaying proton profile that 
covers the mass specified in Table~\ref{abund}.
This method is different from that used by \citet{gallino98}, 
who include a \iso{13}C profile directly 
into the intershell.

\section{Model Results} \label{results} 

\subsection{Germanium to Strontium}

\begin{figure}
\begin{center}
\begin{tabular}{c}
\includegraphics[width=7cm,angle=270]{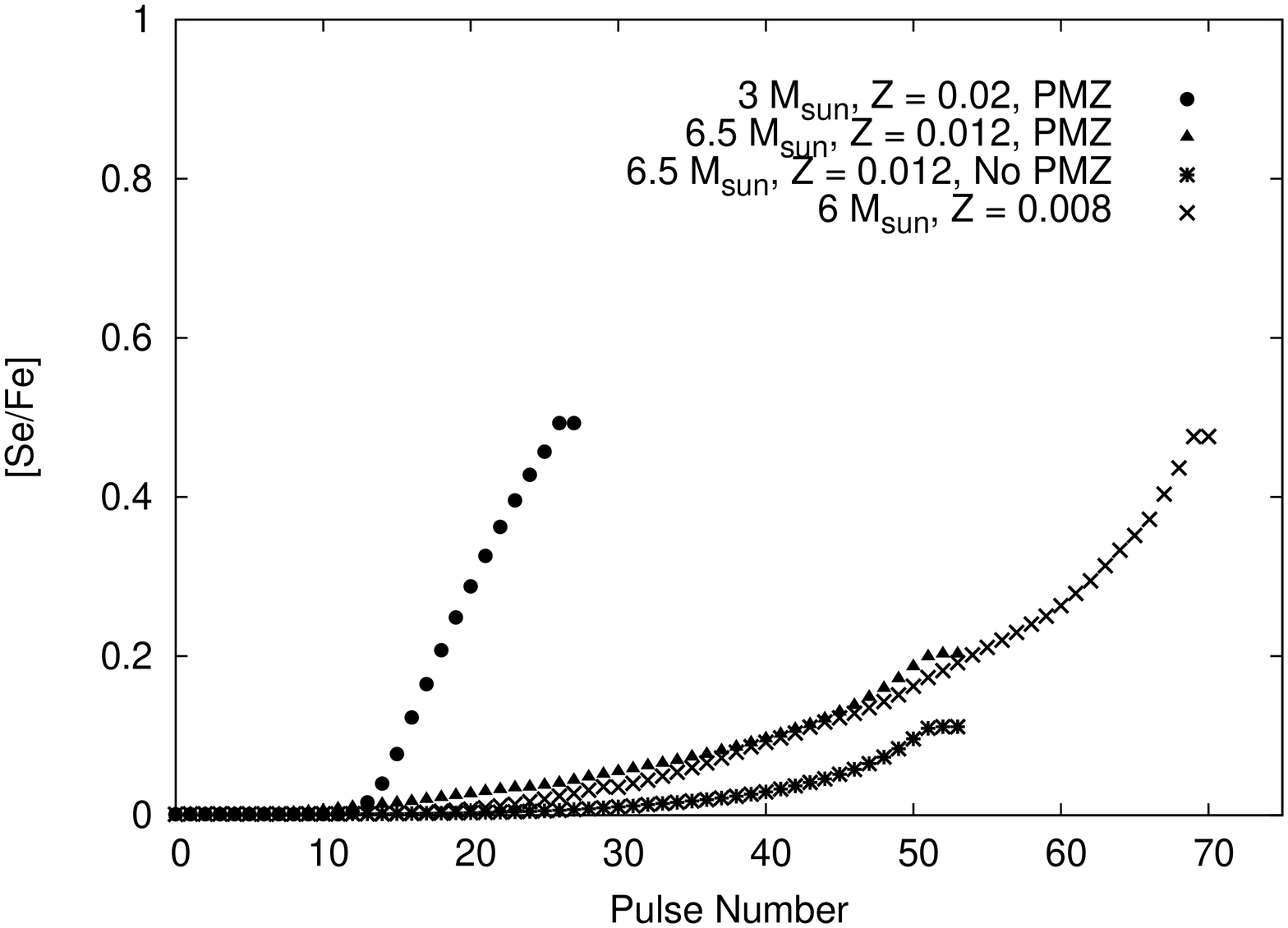} \\
\includegraphics[width=7cm,angle=270]{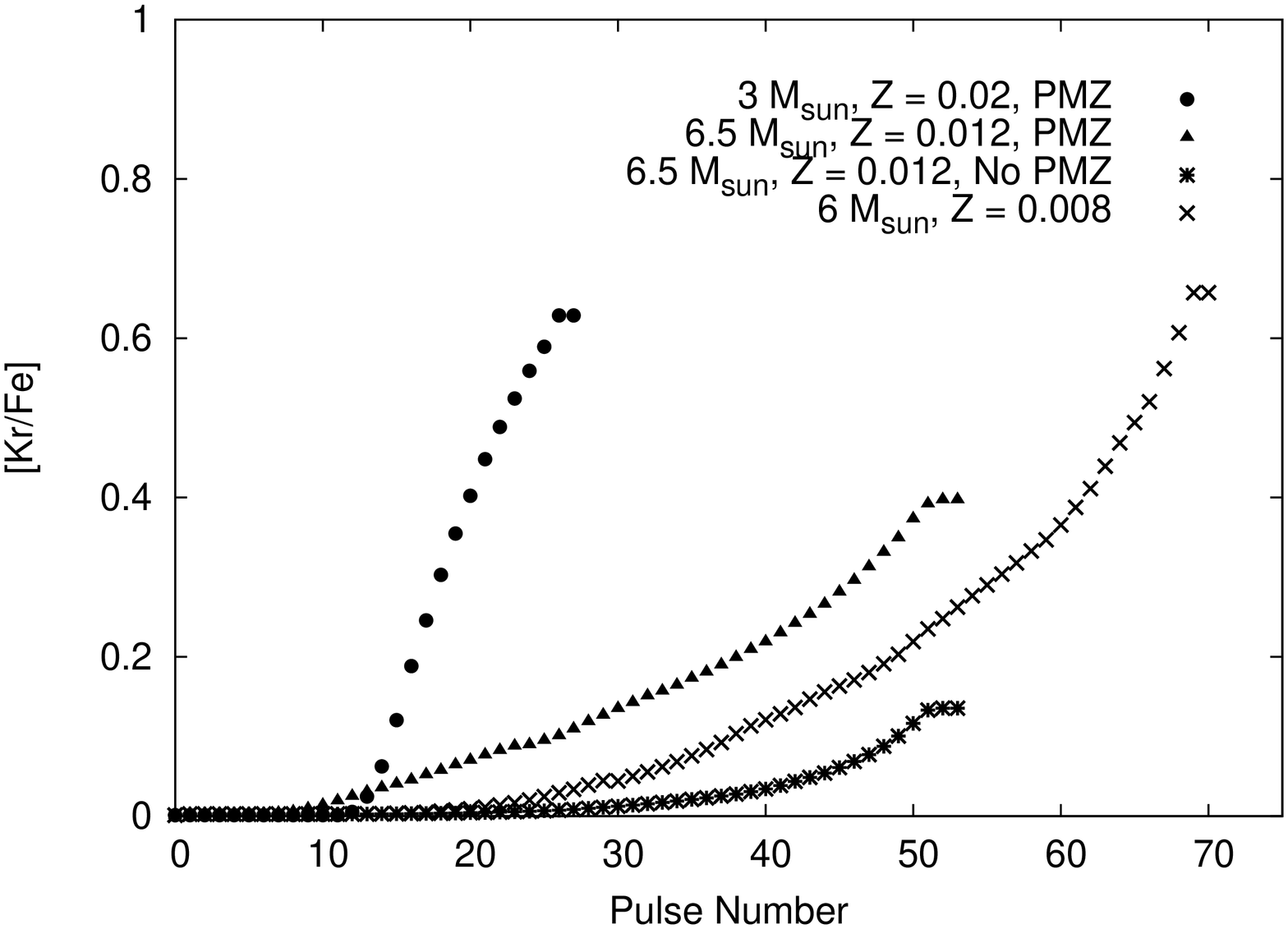} \\
\end{tabular}
\caption{We show the [Se/Fe] ratio (top panel) 
and the [Kr/Fe] ratio (bottom panel), as a function
of the thermal pulse number, from a selected number of stellar
models.  Each point represents the surface abundance during 
the interpulse period.
\label{krplot}}
\end{center}
\end{figure}

Surface abundances at the tip of the TP-AGB for 
Zn to Sr are presented in Table~\ref{abund}, and 
Figure~\ref{krplot} depicts enrichments of Se 
and Kr as a function of pulse number in selected 
models.  At solar metallicity, the 3$\Msun$ models 
with a \iso{13}C pocket produce the largest surface 
enrichments of $s$-process nuclei, with final 
[Se/Fe] and [Sr/Fe] abundances of $\sim 0.50$ and 
$0.85$ for the $Z = 0.02$ model, and 0.56 and 1.2 
for the $Z = 0.012$ model. In the lower metallicity 
models, the largest surface enrichments are seen 
in the 2.5$\Msun$, $Z=0.008$ model with final 
[Se/Fe] and [Sr/Fe] abundances of 0.60 and 1.3, 
and in the 5$\Msun$, $Z=0.004$ model with 
$\sim 0.90$ and 0.93, respectively.
 
The solar metallicity intermediate-mass 
($M \ge 5\Msun$) models exhibit little \emph{s}-process
enrichment ([Kr/Fe] and [Sr/Fe] $<0.25$). Using 
\citet{asplund05} solar abundances, the 6.5$\Msun$, 
$Z=0.012$ model shows slightly larger enhancements 
([Kr/Fe] $=0.40$, [Sr/Fe] $= 0.30$), 
albeit only when a PMZ is added.  
The inclusion of a \iso{13}C pocket has 
little effect ($\lesssim 0.1$ dex) on the production 
of Se and Ge, as noted by \citet{karakas07a}, but 
it does enhance the Kr and Sr abundances.
Only the lower-metallicity 5 and 6$\Msun$ models produce
substantial enrichments of $s$-process elements. These
models experience a large number of TPs ($\gtrsim 60$), 
each with efficient TDU, resulting in significant 
surface enrichments. 
No \iso{13}C pocket was used for these models, and 
thus it is clear that the higher temperatures obtained 
during TPs efficiently activates the \iso{22}Ne neutron
source. Moreover, the number of neutrons per seed nucleus
is larger at lower metallicity owing to the smaller 
abundance of seed nuclei (e.g., \iso{56}Fe).

We find that Kr is enhanced in the stellar envelopes to 
a greater degree than Se, in agreement with the observational 
results of \citet{sterling07} and SD08.  This is because 
Kr lies nearer to the first $s$-process peak than Se, 
and has an isotope (\iso{86}Kr) with a magic number of 
neutrons\footnote{Nuclei with a magic number of neutrons are 
relatively stable against $n$ capture due to their small 
$n$-capture cross sections, and are the cause of the peaks 
in the $s$-process enrichment distribution.}.
Furthermore, the larger $s$-process enhancement 
of Kr relative to Se in the AGB models are in agreement with 
the chemical evolution results 
of \citet{travaglio04}, who find that the main $s$-process 
contributions of AGB stars to the Solar System Se and Kr 
abundances are 14\% and 30\%, respectively (the weak 
$s$-process, operative in massive stars, 
contributes 25\% and 20\% to the solar Se and Kr abundances, 
while the $r$-process provides the remainder).

The [Ge/Fe] abundances in Table~\ref{abund} are different 
from those given in Table~2 in \citet{karakas07a}, owing 
to small changes to the neutron-capture cross sections 
used in the 166 and 156 species networks.  There is a 
0.08~dex change in the [Ge/Fe] abundance for the 3$\Msun$, 
$Z=0.02$ model with PMZ of 0.002$\Msun$. This difference of 
0.08~dex is less than found when considering other model 
uncertainties. For example, there is a 0.16~dex difference 
in the [Ge/Fe] abundances from the two different 5$\Msun$, 
$Z=0.02$ models with different mass-loss laws.

Efficient activation of the \iso{22}Ne neutron source leads 
to positive $\delta$\iso{86}Kr/\iso{82}Kr 
ratios\footnote{The $\delta$ notation is commonly used in meteoritics 
\citep[see, e.g., ][]{lugaro03b}, and is a variation from 
solar in parts permil, or parts per thousand (where a 100 
permil change in the isotopic value is equivalent to a 
10\% variation).}, as can be seen in Table~\ref{abund} 
for the most massive AGB models.  
These positive $\delta$\iso{86}Kr/\iso{82}Kr ratios 
are a result of neutron densities in excess of $10^{8}$ cm$^{-3}$,
which enables neutron captures on \iso{85}Kr that
produce \iso{86}Kr.  If the density is lower, as is the
case when \iso{13}C($\alpha, n$)\iso{16}O is the neutron source,
the branching is bypassed and \iso{85}Rb is produced 
instead, resulting in negative values of the 
$\delta$\iso{86}Kr/\iso{82}Kr ratio. 
The $Z=0.02$ models reflect this trend, showing
an increase in the $\delta$\iso{86}Kr/\iso{82}Kr ratio
with mass, with 5$\Msun$ the minimum mass for efficient
activation of the \iso{22}Ne source.
The ratio is negative in the 3--4$\Msun$ models,
where the \iso{13}C reaction dominates as the neutron 
source and the \iso{22}Ne reaction is only marginally 
activated.
Values of $\delta$(\iso{86}Kr/\iso{82}Kr) measured in 
meteoritic stellar SiC grains from AGB stars range 
roughly from $-$500 to 900, and increase with the grain 
size \citep{lewis94}. Negative values are well explained 
by AGB model predictions, however, it is difficult to 
explain the large observed positive values because 
they are only produced by HBB models, where the C$>$O 
condition necessary for the formation of SiC grains is 
not attained, except in the lowest metallicity, $Z=0.004$
models. Further detailed work is required on this issue.

The resulting elemental abundance pattern also
differs, as Sr is produced in favor of Rb in the 
low-neutron density case, resulting in low Rb/Sr ratios
($\sim 0.05$). This is consistent with observations of 
low-mass AGB stars \citep{lambert95,abia01}. 
In the high-neutron density case, Rb is produced instead,
as observed in massive O-rich AGB stars \citep{garcia06}.
Preliminary results for Zr and Rb from intermediate-mass
AGB stars are discussed in \citet{vanraai08}, 
and will be presented in more detail in a forth-coming 
paper (M. A. van Raai et al., in preparation).

\subsection{The iron-peak elements Iron and Zinc}

In Table~\ref{abund}, we show the [Zn/Fe] ratios and
$\Delta$Fe values for each model, where $\Delta$Fe is the logarithmic
change to the surface iron abundance from the main sequence
to the tip of the TP-AGB.  The $\Delta$Fe values are all very 
nearly zero, indicating that the surface abundance of this 
element is unaltered by AGB nucleosynthesis.
The maximum effect among the $Z$ = 0.02 models is 2\%, 
and even for the $Z$ = 0.004 case the Fe abundance decreases
by less than 5\%. This makes Fe a suitable reference element 
for estimating metallicities in AGB stars.

% initial Fe 56, 57, 58 = 2.33142e-05  1.10506e-06  1.31483e-07 => total = 2.4550743E-05
% final intershell      = 2.52587e-06  1.33822e-06  6.01659e-06 => total 9.8806800E-06 is 40% of initial

Although the surface Fe abundance is not significantly 
changed by nucleosynthesis during the AGB, the abundance
of Fe in the He-intershell can be depleted by up to
$\sim 1$~dex in a massive AGB star (e.g., 6.5~$\Msun$ 
model). In the intershell \iso{56}Fe is destroyed 
by neutron captures, and as the dominant isotope it 
primarily determines the elemental abundance (even if 
the abundances of the rare neutron-rich isotopes \iso{57}Fe 
and \iso{58}Fe increase). 
In lower mass models (e.g., 3~$\Msun$), depletion of Fe 
is restricted to the \iso{13}C pocket that encompasses 
10 to 15\% of the mass of the intershell. The next TP 
homogenizes this small, Fe-depleted region
with the much larger Fe abundance in the rest of the 
shell.  Our 3~$\Msun$ results are consistent with those
of \citet{werner06}, who found that AGB models cannot 
account for the large Fe depletions found in some central 
stars of PNe, particularly the hot, H-deficient stars of 
the PG1159 class \citep{miksa02,rauch08}.
\citet{werner06} estimated that these central stars
have mean masses of $\sim 0.62\Msun$, and thus represent
the evolutionary remnants of lower-mass progenitors than
those addressed here, and should be compared with 
models of stars with $\lesssim 3\Msun$.
Note that similar or even more extreme
Fe deficiencies -- along with dramatic overabundances of
$n$-capture elements -- are seen in older
white dwarfs where mechanisms such as gravitational
settling and levitation by selective radiation pressure
play important roles in establishing the
surface abundances \citep{chayer05,werner06}.

In PNe, Fe can be depleted into dust by
large factors that may vary from object to object 
\citep{perinotto99}, and even in different regions within 
the same nebula \citep{sterling05a}, so that its gas-phase
abundance is not a reliable indicator of the true elemental
abundance.
The heavier iron-group element Zn exhibits much milder
depletions in the ISM, particularly in the warm phases
\citep[e.g.,][]{welty99}.
Consequently, Zn is frequently used as a proxy for iron and the
iron-group, for example in damped Lyman-$\alpha$
absorbers \citep[e.g.,][]{akerman05}.
However, the use of Zn as an indicator of the initial
composition of an AGB star or PN requires that the change
in the abundance of Zn due to internal nucleosynthesis
is small compared to its initial abundance in the star.
It can be seen from Table~\ref{abund} that the predicted
changes in the Zn abundance are modest or negligible 
for most of the models.
This is consistent with the conclusions of 
\citet{travaglio04}, who estimated the total contribution 
of AGB stars to the solar Zn to be $\sim 3$\%, and with
\citet{cayrel04}'s finding that Zn production in the 
Milky Way is dominated by other sources (i.e., 
explosive Si burning in massive
stars).  However, at low metallicities, the models
show a large enough increase that the Zn enrichment
may be detectable.

These results can be compared to the Zn abundances derived
for two Type~I PNe (DGS08).  Of the models presented here, 
the Galactic disk PN M~1-40 is most closely matched by the 
6.0 $M_{\sun}$ model with $Z$ = 0.02, given its solar Ne 
and S abundances and N/O~=~1.13 \citep{gorny04}.  This model
predicts a very small enrichment of Zn, [Zn/Fe]~$\sim0.038$, 
which is consistent with the measured value (see \S~2), 
even allowing for a possible small 
depletion effect. The case of LMC~SMP~62 is more problematical.
According to its $\alpha$-element abundances, the most
appropriate models are those for $Z$ = 0.008. At this 
metallicity, all three of the models in Table~\ref{abund}, 
calculated for 2.5, 5.0, and 6.0 $M_{\sun}$ respectively, 
predict similar Zn enhancements of $\sim$0.15--0.2~dex.  
However, as described in \S~2, the measured Zn abundance in 
this PN is lower, rather than higher, than those of the 
$\alpha$-elements (e.g., Ne, S), in contradiction to the models.
The disagreement is even worse for Se, which is predicted to be
enhanced by 0.4--0.6~dex, depending on the initial mass.  
A modest enrichment in both Zn and Se is possible, if 
their initial abundances were low enough 
relative to Fe, but nevertheless it remains difficult
to satisfy the observational constraints for LMC SMP 62 with the
listed models. We note, however, that these particular models 
also predict N/O values either lower (the 2.5 $\Msun$ model) 
or higher (the 5 and 6 $\Msun$ models)
than the observed value for this PN; in \S~5 we discuss other 
possible scenarios for explaining the observed properties of 
LMC SMP 62.

\subsection{Neon through to Argon}

Abundances of S, Cl, and Ar in  PNe are
powerful tools for studying the chemical evolution of 
galaxies and stellar systems, under the assumptions that 
these elements are unaltered by AGB evolution 
\citep{dopita97,stasinska98,leisy06,idiart07} and are 
produced in the same relative proportions in massive stars 
\citep{stasinska98}. The case of Ne is not as clear, 
although the remarkably constant Ne/O ratio of $\sim$~0.20 
(by mass) observed in PNe with a wide variety of chemical 
compositions indicates that in most cases this ratio is 
not altered during the AGB. ~\iso{20}Ne, the most abundant 
isotope of Ne, is not altered significantly by H or He-burning 
in AGB stars, whereas $^{22}$Ne can be produced by successive 
$\alpha$-captures onto $^{14}$N.
Therefore an enhancement in the Ne/O ratio is expected only 
when \iso{22}Ne is sufficiently enriched that its abundance is 
comparable to or exceeds that of \iso{20}Ne.  \citet{karakas03b}
predicted that such an increase would only occur in a narrow 
range of progenitor masses, $\sim 2.5$ to $\sim 3.5\Msun$ 
depending on metallicity.  Our current models are generally 
consistent 
with these earlier results, although from Table~\ref{abund-2} 
it is clear that the 6.5~$\Msun$, $Z = 0.012$ model and the 
5 and 6~$\Msun$, $Z= 0.008$ and 0.004 models have enhanced 
Ne/O ratios as a result of some O destruction by HBB, and 
mild enrichments of Ne via dredge-up 
(maximum Ne increase of $\sim 0.3$~dex for the 5, 6$\Msun$
metal-poor models). Therefore, Type I PNe are not expected 
to be significantly enriched in Ne except at low metallicities.

The \iso{22}Ne is produced primarily in the TPs themselves, 
but in lower mass models (2.5 and 3$\Msun$), the
partial mixing of protons can further enhance the 
Ne/O ratio.  This is because the partially
mixed zone results in a pocket of \iso{13}C and \iso{14}N,
thus increasing the \iso{14}N abundance in the intershell
to values higher than that from CN-cycling
during the interpulse. The next TP quickly converts
the \iso{14}N into \iso{22}Ne. The resulting \iso{22}Ne 
abundances after the 
TP are slightly higher than in the case with no 
partially-mixed zone.  After many TPs and TDU, the PMZ 
results in an increase in the elemental Ne abundance. 
The observations by \citet{bernard08} are intriguing
in this respect.
They determined Ne/S ratios in a sample of 25 PNe in the 
Magellanic Clouds, and found that four of the objects show 
high Ne/S ratios. For two of the objects, they attribute the
high ratios to low S abundances,  but in the other two PNe, 
the high Ne/S values could be due to Ne enrichments.

Table~\ref{abund-2} shows the surface abundance predictions 
for S, Cl, and  Ar. These elements generally are not altered
by AGB nucleosynthesis processes by more than the observational
errors ($\sim \pm 0.1-0.15$~dex), consistent with the 
observations of PNe by \citet{stasinska98}. Neutron captures
in the He-shell can result in variations in the isotopic ratios
involving the stable Cl and Ar isotopes; this is demonstrated
in the last column of Table~\ref{abund-2} where we show 
the permil variations for \iso{38}Ar/\iso{36}Ar. Values 
of $\delta$(\iso{38}Ar/\iso{36}Ar) measured in stellar 
SiC grains range from zero (solar composition) to 
positive values up to $\sim 200$ \citep{lewis94}, in 
qualitative agreement with our model predictions.  
\citet{kahane00} compared AGB model predictions to 
measurements of Cl, Mg, Si, and S isotopic ratios in
the cool carbon star IRC$+$10216, and inferred an 
initial mass of $\lesssim 2\Msun$ based on the accuracy of
the Cl and Mg data. Kahane et al. concluded that a larger mass 
progenitor would lead to too high Cl and Mg isotopic 
ratios at the surface, although our models suggest that
progenitor masses as high as $\sim 3\Msun$ produce a
good fit to the observed data, within the errors
quoted for the isotopic ratios.
In the case of our low-metallicity massive AGB 
models, neutron captures result in modest increases 
($\sim 0.25$~dex) in the elemental Cl surface 
abundance.

The phosphorous abundance can be determined in hot post-AGB 
objects such as the PG 1159 stars, whose envelopes are composed 
of He-intershell material. Observations indicate that
PG 1159 stars have solar P abundances, whereas
3$\Msun$ models predict P abundances ranging from four times 
solar to $\sim 25$ times solar in the He-intershell, 
depending on the amount of extra mixing included in the 
calculations to produce the $^{13}$C pocket \citep{werner06}. 
Our He-shell predictions
are consistent with Herwig's model, in that we estimate P 
enhancements from 1.3 to $\sim 9$ times solar for the 
3~$\Msun$ model, without and with a PMZ respectively.
The large P abundances in the He-intershell abundances do not 
result in large surface enrichments ($\lesssim 0.1$~dex
in all solar metallicity models), so that if measurements 
could be made of the P abundances in PNe, we would not 
expect to see variations from solar.

Sulfur is also observed in PG 1159 stars, with abundances varying 
from 0.01 times solar to solar \citep{werner06}.  
It is not clear if this range reflects the initial 
abundance of the star, various degrees of nuclear processing, 
condensation of S into dust or problems in determining 
accurate S abundances.
Note that S abundances in several Galactic \citep{henry04} 
and some Magellanic Cloud PNe \citep{bernard08} are low 
relative to O.  While infrared observations of S$^{3+}$ 
in Galactic PNe seem to indicate that 
uncertainties in corrections for unobserved ionization 
stages are not the reason for the low S abundances, it 
is possible that S may be depleted into dust in some 
PNe \citep[especially C-rich objects; ][]{pottasch06}.
AGB models do not predict large
depletions of S in the He-shell, with S abundances 
varying between 0.6 to 0.9 times solar \citep{werner06}.
Our values lie from no change to $\sim 0.9$ times solar.
The results for P and S are puzzling, indicating
on the one hand that a smaller PMZ is required 
(to produce less P), or, on the other hand that
more mixing is required (to destroy more S). 

\subsection{Oxygen in intermediate-mass AGB stars} 

Oxygen constitutes a large fraction of the total metal 
content of a star, and in this sense is a more fundamental 
measure of metallicity than Fe/H, which historically 
has been used as the primary indicator of metallicity 
due to its ease of measurement. While it can be 
challenging to determine O abundances in stars 
\citep[e.g.,][]{ramirez07}, O/H is easily 
measured in ionized nebulae because of its
bright emission lines, and the resulting 
abundances are considered to be among the most accurate for 
any element. According to the current paradigm for galactic 
chemical evolution, O production is dominated by massive 
stars ($M~\geq~10~\Msun$). Nucleosynthesis models of AGB 
stars indicate that, collectively, these stars do not 
contribute significantly to the production of O
\citep{forestini97,karakas07b},  although there is 
evidence for dredge-up of self-produced oxygen in a 
few special cases, including PNe from metal-poor 
progenitor stars \citep[e.g.,][]{pequignot00,dinerstein03}.
The inclusion of diffusive convective overshoot 
at the inner edge of
the flash-driven convective zone during a TP can
lead to O abundances in the intershell of $\sim$~20\%
\citep{herwig00}, in contrast to standard models
that predict 2\% by mass or less \citep{boothroyd88c,karakas02}. 
This can lead to increases in the surface composition 
of O, particularly at low metallicities.
Evidence for this overshoot comes from PG 1159 stars 
that show compositions consistent with Herwig's models 
\citep{werner06}. It is unclear however, if this 
overshooting occurs for all AGB stars or only for AGB 
stars that produce PG~1159 stars as a consequence of 
late and very late TPs. For example, the observed 
abundance analyses in intrinsic and extrinsic AGB stars,
as well as in SiC grains, suggest that such an overshoot
is uncommon \citep{lugaro03a}. This is because 
in the region of the He-rich intershell where primary O 
is synthesized, the temperature is large enough for 
efficient activation of the \iso{22}Ne($\alpha,n$)\iso{25}Mg
neutron source, even in low-mass AGB stars. 
In this case, the resulting neutron density is very 
large (up to $\sim 10^{12}$ neutrons/cm$^{3}$) in
contrast to observations of low-mass AGB stars
that indicates neutron densities of about 
$\sim 10^{7}$ neutrons/cm$^{3}$ \citep{abia01}.

Intermediate-mass AGB stars on the other hand, 
destroy O via HBB when the temperature 
at the base of the convective envelope is sufficiently 
high ($T \gtrsim 80 \times 10^{6}$~K).  The O-Na 
anti-correlation observed in globular clusters stars 
has in fact been associated with HBB in 
intermediate-mass AGB stars, because HBB can 
qualitatively account for the destruction of O 
along with the production of Na. Moreover, SD08 noted 
that the Ar/O, S/O, and Cl/O ratios of Type~I PNe 
are approximately a factor of two larger on average 
than in non-Type~I objects. Since Ar, S, and Cl 
are not significantly affected by nucleosynthesis in
low- or intermediate-mass stars (see \S~4.3), they 
interpreted this as evidence for O destruction during 
HBB in Type~I PN progenitors. This result had also been 
found in previous investigations 
\citep{marigo03,leisy06,pottasch06}.

The amount of O destroyed by HBB depends
on the temperature at the base of the convective envelope.
The temperature, in turn, depends on 
a number of factors including the initial mass and metallicity,
with hotter temperatures found in models
of increasing mass for a given $Z$, or in models of 
decreasing metallicity, at a given mass. In addition, 
the predicted efficiency 
of HBB is sensitive to the treatment of convection and 
mass loss in the models, neither of which are 
well-constrained empirically.
In particular, the convective efficiency affects 
the luminosity and the mass-loss rate \citep{ventura05a}, 
and hence these properties are sensitive to the convective 
model that is utilized.  Mass loss determines the 
length of the HBB phase, since HBB does not operate 
once the envelope mass drops below about 1.5 $M_{\odot}$.  

In Table~\ref{abund-2} we present the O abundances
at the surface of the model star at the tip of the AGB. 
The O abundance remains unaltered in the lower mass 
models (3 and 4$\Msun$, $Z = 0.02$), whereas
O destruction takes place in models with HBB.
Out of the $Z=0.02$ models, the largest amount of O 
destruction is seen at 6.5$\Msun$ which has
a peak temperature at the base of the envelope of 
86$\times 10^{6}$K.  The effect of metallicity is seen
most clearly by comparing the 6.5$\Msun$ models
with $Z=0.02$ and $Z=0.012$, with the latter model 
computed assuming \citet{asplund05} C, N, O initial 
abundances. In the $Z=0.012$ model the peak temperature 
is higher, resulting in 0.15~dex destruction of O 
compared to only 0.09~dex in the $Z=0.02$ model. 
Note that the O that is destroyed is all converted 
into N. The largest O destruction of 0.26~dex occurs 
in the 6$\Msun$, $Z=0.008$ model. Overall, none of 
the models suffered O destruction of more than 0.3~dex.

It is not clear that the low O abundances relative to
S, Ar, and Cl in Type I PNe can be attributed to HBB.  While 
modeling uncertainties in the treatment of convection and 
mass-loss could in principle allow for more efficient O 
destruction during HBB than we report, this inevitably leads 
to large increases in the N/O ratio.  This result agrees with 
the findings of \citet{marigo03}, whose synthetic models with 
efficient ON-cycling produced N/O ratios considerably higher 
than those observed in Type~I PNe.  Those authors suggested 
that the low O abundances may be explained if the initial 
metallicities of Type~I PN progenitors are low.  However, 
this does not explain the approximately solar abundances of 
S and Ar in these objects.  Moreover, our models constrain 
the metallicities of Type~I PN progenitors to be 
$Z\gtrsim 0.4Z_{\odot}$ in order to reproduce the observed 
Se and Kr abundances.  
This important issue remains unresolved.

\section{Discussion} \label{discussion}

\begin{figure}
\includegraphics[width=8cm,angle=270]{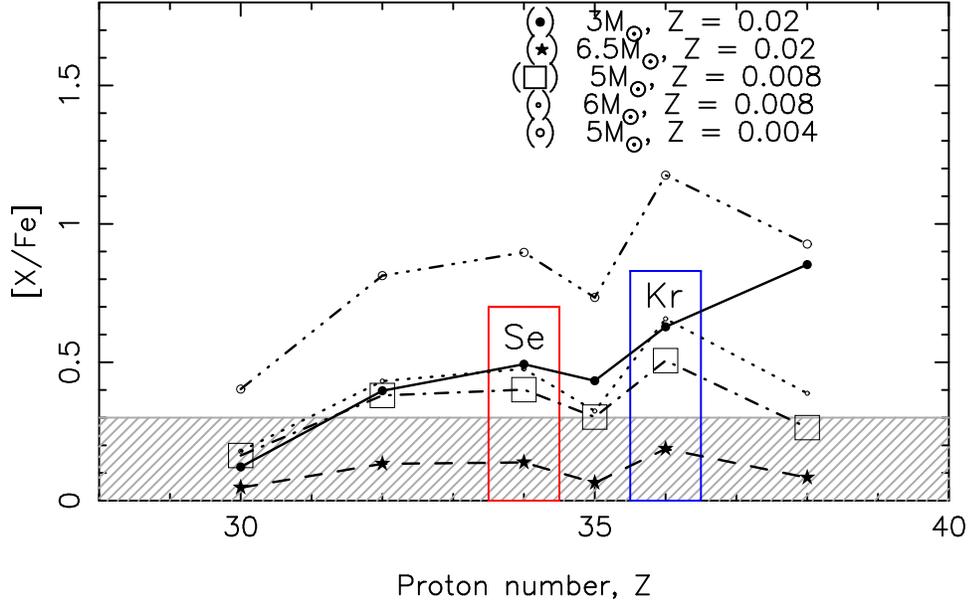}
\figcaption{The surface abundances of Zn, Ge, Se, Br, Kr,
and Sr at the tip of the AGB phase for a selection of the 
stellar models. Abundances are shown as [X/Fe] ratios,
and are plotted as a function of proton number, $Z$.
The 3$\Msun$, $Z=0.02$ model includes a PMZ of 0.002$\Msun$ 
whereas all other models are without \iso{13}C pockets.
The gray-hatched box with a maximum at [X/Fe] = 0.3 represents
the region of observed Type I PN Se and Kr abundances. 
Note that observational uncertainties allow for [Se, Kr/Fe] 
up to 0.6.  Boxes around
the elements Se and Kr indicate that for the present study we
can only use these elements for comparison as large PN abundance
data-sets of the other trans-iron elements are not yet available.
\label{sabunds}}
\end{figure}

In Fig.~\ref{sabunds} we illustrate the $s$-process 
abundance predictions from a selection of the stellar models.
The gray-hatched box represents the approximate region 
occupied by Type I PNe $s$-process abundances. 
Typical observational uncertainties allow the maximum 
enrichment to reach $\approx0.6$ dex in some cases.
The intermediate-mass AGB models are, 
in general, a good match to the observed abundances of 
Se and Kr in Type I PNe. Models in Table~\ref{abund} 
that show enhancements of Se and Kr 
greater than 0.5~dex are the 2.5$\Msun$ model, the 3$\Msun$ 
models, the  6$\Msun$, $Z = 0.008$ model, and the 
low-metallicity 5$\Msun$, $Z = 0.004$ model, which
gives large enhancements for all elements between Zn and Sr. 
The results for 3$\Msun$ with a PMZ are only 
in borderline agreement with the observations, 
although the results with no PMZ are consistent with
no $s$-process enrichments. In either case
the final composition is carbon-rich, indicating
that this model is unlikely to be the precursor of 
a Type I PN.
Indeed, the results for the 3$\Msun$ models are
closer to that found for the 2.5$\Msun$, $Z=0.008$
model which has a distinct, non-Type~I PNe composition,
with a high final C/O ratio, and considerable 
overabundances of all $s$-process elements 
($>$~0.5~dex for Se to Sr).
The results for the 5$\Msun$, $Z = 0.004$ model 
suggests that Type I PNe in the Galaxy do 
not descend from low-metallicity stars, 
as suggested, for example, by \citet{marigo03}, 
and that a minimum metallicity of $\sim 0.4~Z_{\odot}$ 
is required for these PNe. 
This result is not unexpected, given that Type I PNe 
evolve from more massive stars than other PNe, and 
hence trace a younger population. On the other hand, 
these model results suggest that Type I PNe in the
Magellanic Clouds could be more highly enriched in 
light \emph{n}-capture elements such as Se and Kr 
than their Galactic counterparts.
This is a strong motivation for further observations 
of $s$-process elements in PNe belonging to the Magellanic 
Clouds or to other metal-poor populations.

Observations of other trans-iron elements in PNe in 
addition to Se and Kr could be used to constrain the mass 
and metallicity ranges of Type I PNe.  For example, Sr 
and Rb abundances are potentially useful discriminants 
of progenitor mass.  Sr is considerably enriched in 
the $3M_{\odot}$ models ([Sr/Fe]~$\sim0.85-1.23$), but 
is slightly 
if at all enhanced in the 6 and $6.5 M_{\odot}$ models 
([Sr/Fe]~$<0.4$ dex). Note that the inclusion of a PMZ in the
6.5$\Msun$, $Z=0.012$ model increases the level of Sr
enrichment expected, from 0.05 to 0.3~dex.

Of these next two elements heavier than Kr (element 
number 36), Rb (element 37) is more likely to be 
measurable in a meaningful way in nebulae than Sr 
(element 38), for the following reasons. 
First, Sr is refractory and can be strongly depleted 
into dust grains. In fact, based on its condensation 
temperature \citep{lodders03}, Sr is expected to be 
incorporated into dust to a greater degree than Fe, 
whose gaseous abundance is depleted by 1--2 orders of 
magnitude in PNe \citep{perinotto99,sterling05a} and 
the ISM \citep{welty99}.  Therefore, measurements of 
the gaseous Sr abundance in PNe are not likely to reveal 
useful information about \emph{s}-process enrichments.  
Second, the only emission lines of Sr in the optical 
spectral region originate from relatively highly ionized 
states that are not expected to be abundant, except in
a few of the highest-excitation PNe. Sr emission has been 
identified in only one PN to date 
\citep{pequignot94, zhang05}. 
On the other hand, based on its condensation temperature 
\citep{lodders03} Rb does not deplete into dust as 
readily as Sr, and thus its gaseous 
abundance should be more representative of the overall 
elemental abundance. Rb emission has been identified
in a number of PNe \citep{pequignot94, zhang05,sharpee07} 
and should be detectable in PNe with a broader range of 
ionization than Sr.  This discussion suggests that Rb 
is a useful element with which to constrain stellar models 
to PN observations.

Furthermore, Rb is observed to be substantially enhanced 
in massive Galactic OH/IR stars with the maximum 
enhancement [Rb/Fe] $\sim 2 \pm 1$~dex  \citep{garcia06}.
\citet{vanraai08} have made a preliminary comparison 
between  Zr, Rb, and Li predictions from 5 to 
6.5$\Msun$ solar composition models to the observations
of the OH/IR stars.
The best match between stellar models and the 
observations was obtained with the 6.5$\Msun$
model computed by 1) not including a 
\iso{13}C pocket, and 2) by accounting for extra 
TPs not modeled in detail, as done by \citet{karakas07a}. 
The extra TPs and TDU mixing episodes may occur
because not all the envelope mass was lost when 
convergence difficulties ended the computation. 
The main uncertainty is the behavior of the TDU 
efficiency at small envelope masses, with 
studies generally finding a decrease in the 
efficiency with decreasing envelope mass
\citep{straniero97,karakas02}.  Further studies 
on the evolution of the TDU efficiency with 
decreasing envelope mass are needed to help settle 
this issue.  The tabulated results presented 
in this study were calculated without the inclusion of 
these extra TPs and can be considered lower limits to 
the final PN abundances, noting that it is highly 
uncertain how much of an impact the final TPs would 
have on the the final PN abundances.

The final few TPs can potentially have a significant effect 
on the final abundance of the star and its resulting PN.
For example, \citet{frost98a} noted that intermediate-mass 
AGB stars may become luminous, optically obscured carbon 
stars near the end of the TP-AGB, when mass loss has 
removed much of the envelope, extinguishing HBB but 
allowing dredge-up to continue. Under these conditions
the envelope mass is signficantly reduced, resulting in
only minimal dilution of the intershell material that
is dredged up. This can allow for large enhancements to the
surface composition of C and other He-burning 
products. In the models with HBB included in this study,
all experienced a few TPs and TDU episodes after the 
cessation of HBB, with some of the lower metallicity models 
becoming carbon rich (see Table~\ref{stellarmodels}, 
i.e., 6$\Msun$, $Z=0.008$; 5 and 6$\Msun$, $Z=0.004$) 
before convergence difficulties ended the computation. 
Note that it is after this stage that we consider 
the effect of remaining TPs not modeled in detail 
using a synthetic AGB algorithm.

In \citet{vanraai08} the inclusion of these remaining
TPs only had a large impact on the surface abundances
of the 6.5$\Msun$, $Z=0.02$ model (out of 
the 5, 6 and 6.5$\Msun$, $Z=0.02$ models),
and only when assuming that efficient TDU occurs
at a very small envelope mass\footnote{Besides the
inclusion of the extra TP the models presented 
here and in \citet{vanraai08} are the same.}.
A final [Rb/Fe] $\approx 0.9$ was obtained 
in the 6.5$\Msun$, $Z=0.02$ model, in agreement with 
the observed Rb abundances of OH/IR stars given 
the large uncertainties. Along with increases in Rb, 
the [Se, Kr/Fe] abundances were increased to 0.6 and 
0.7~dex (respectively), above the observed 
Type I PN limit of 0.3~dex. 
This model was estimated to have 7 remaining TPs,
with the last TP occurring with a very small 
envelope mass of $\sim 0.03\Msun$. The small
envelope mass during the last TDU leads to the 
situation discussed above: A substantial increase 
in the surface abundances, caused by the mass of 
the envelope being only $\sim 10$ times greater 
than the mass of the intershell.  It is questionable 
whether such an efficient TDU would occur for 
such a low envelope mass; so if we exclude this 
last mixing episode, the [Se, Kr/Fe] abundances are 
$\approx 0.4$ and 0.45, respectively, in agreement 
with observed Type~I PN abundances.
This analysis may indicate that the OH/IR stars 
are {\em not} the progenitors of most Type I PNe 
and that there are possible observational biases 
in the sampling of each population.
That is, the OH/IR population may sample the
most massive AGB stars (6--7$\Msun$), with the
largest [Rb/Fe] (and [Kr/Fe]) abundances, while
the Type~I PN population may come from slightly less
massive AGB stars, with lower Kr and Rb abundances.
There is also the problem of explaining the
large Rb abundances in the OH/IR stars, given that
they seem to require the TDU to operate efficiently 
at small envelope masses. 
More observations of Rb in OH/IR stars and PNe
will help to settle these issues. 

Modeling uncertainties can dramatically affect AGB
nucleosynthesis predictions. As noted previously, 
different convective models have a large impact on the
nucleosynthesis of light elements (e.g., N, O, Na) in AGB
models with HBB. Improvements in our understanding 
of convection are desperately needed but this 
may only come about through  multidimensional studies 
\citep[e.g.,][]{dearborn06,herwig06,meakin07}. 
The mass-loss law used in the computations is similarly 
important, because this can determine (for a fixed 
convective model) the number of TPs and mixing episodes. 
For example, if we artificially cut the number of TPs 
by half for the 5$\Msun$, $Z = 0.004$ model (40 TPs 
instead of the computed 80), the final [Kr/Fe] ratio 
is $\sim 0.35$~dex instead of 1.17~dex. This simple
analysis indicates that we need better constraints
on the mass loss from low-metallicity, massive AGB stars
if we are to provide constraints on the metallicity
range of Type I PNe.   

Another important uncertainty relates to the unknown 
formation mechanism for \iso{13}C pockets in
AGB stars \citep{gallino98,herwig05}. These pockets 
are necessary to provide neutrons in low-mass stars. 
It is unclear, however, what role the \iso{13}C
neutron source plays in more massive AGB stars. 
We had hoped to ascertain the dominant neutron source 
from comparisons of Type I PN abundances to massive AGB
stellar models, but the Se and Kr abundances
alone are not sufficient. Selenium abundances were
not strongly affected by the inclusion of a \iso{13}C
pocket in the 5 and 6.5$\Msun$ models, whereas the 
abundance of Kr was affected but not increased above
0.5~dex. Note that in three Type I PNe of SD08
where both Se and Kr were detected, an abundance
of [Kr/Se] $=0.5$ was found, providing some
support for the inclusion of a \iso{13}C pocket
in massive AGB models; this conclusion is, however,
limited by small number statistics.
We predict that PNe should have larger enrichments
of Kr than Se. While this result is valid regardless 
of whether or not a $^{13}$C pocket is included, 
larger [Kr/Se] values are predicted in models with a 
PMZ (see Table~2). Firmer conclusions regarding the 
neutron source operating
in intermediate-mass AGB stars may come from
large data sets of many $n$-capture elements, 
including Br, Rb, and Xe.

Finally, models with extra mixing may be 
necessary to reproduce the observed abundances of 
some PNe. For example, the composition determined for
LMC SMP~62 (see \S2 and \S4.2) suggests that it is a 
Type~I PN, but its low Se and Zn abundances are at odds
with our predicted abundances for intermediate-mass models.  
A $3 \Msun$, $Z=0.02$ model with rotation by 
Cantiello \& Langer (private communication) had a N/O ratio 
of 0.43 after the first dredge up, compared to 0.32 from 
our model with no rotation, whereas the N/O value of 
LMC SMP 62 is $\sim$ 0.4 -- 0.5.
During core H-burning, rotation mixes processed 
material to layers further out than they would otherwise 
be found.  Rapid rotation may 
also inhibit the production of $s$-process elements during 
the AGB phase \citep{herwig03}, consistent with the lack of 
Zn and Se enrichments found in this object. An alternative 
is that LMC SMP 62 evolved from a low-mass progenitor of 
less than $2 \Msun$, which did not experience efficient TDU. 
Note that the low C/O ratio of 0.17 found for this object 
\citep{aller87} is also consistent with a star that did 
not experience any TDU mixing (suggesting an initial 
mass of $\lesssim 1.25\Msun$). It is unclear if rotation 
can explain the high N/O ratio in such a low-mass star. 
Other extra-mixing processes (e.g., thermohaline 
mixing) may need to be invoked in this case.

\section{Conclusions} \label{conclusions}  

Type I PNe have high He/H and N/O ratios, indicating 
that they are the descendants of intermediate-mass 
AGB stars with initial masses between $\sim 3$ to 8$\Msun$. 
\citet{sterling08} found that Type I PNe exhibit 
significantly smaller enrichments of Se and Kr 
($\lesssim 0.3$~dex) on average than other PNe.  
We calculated $s$-process enrichments in a set of AGB models
covering a range in mass from 2.5 to 6.5$\Msun$, and 
metallicity from 0.2$Z_{\odot}$ to $Z_{\odot}$.
The 2.5$\Msun$, $Z=0.008$ model was included to show an 
example of a model that would produce a clear non-Type~I 
PN abundance signature, with 
[Se, Kr/Fe] $>0.5$ and C/O $\sim 4$.
The main conclusion is that the results for the 3--6.5$\Msun$ 
are a good match to the observed abundances. 
The only real exception out of the HBB models is the 
low-metallicity
5$\Msun$, $Z = 0.004$ model that produced much larger 
enhancements in Se and Kr than observed. This suggests
that Galactic Type I PNe do not descend from such 
low-metallicity objects.

We also compare calculated abundances for selected
intermediate-mass (Ne, P, S, Ar) and iron-peak (Fe, Zn) 
elements to observations of post-AGB stars and PNe.
We find that the elemental abundances of P, S, Cl, Ar, 
Fe, and Zn are essentially unchanged by AGB nucleosynthesis, 
although there are isotopic shifts caused by neutron captures 
in the He-shell. These results justify use of elements such 
as S, Cl, and Ar in PNe as tracers of Galactic chemical 
evolution.

It is difficult to reach firm conclusions
about the neutron source operating in massive AGB stars
from Se and Kr abundances in Type I PNe.
Certainly it seems that at the least the \iso{22}Ne source, 
with efficient TDU, is required to produce enhancements in 
$s$-process elements in massive Type I PN progenitors. 
It is less clear whether a \iso{13}C pocket is also 
required in these stars. Increases in
the [Kr/Fe] abundance ratio were observed in the models 
when a \iso{13}C pocket was included,  but not
beyond the amounts observed in Type~I PN spectra.
Obtaining abundances for more $n$-capture elements, 
particularly Rb and elements beyond the first $s$-process 
peak, from a large data set of OH/IR stars and PNe will
help distinguish among the possibilities.

Finally, only the models with HBB ($M \gtrsim 5\Msun$, 
depending on $Z$) show the high He/H and N/O ratios that
define the Type I PN class.  Given that $3 \Msun$ stars 
are more common than $6.5 \Msun$ stars (according to
initial mass function and evolutionary time 
considerations), it may be necessary that another 
mixing process other than HBB is active in 
stars of the lower-mass range (3--$4\Msun$), if 
these stars do in fact evolve into Type~I PNe. Rapid 
stellar rotation in single $\sim 3\Msun$ stars may, 
for example, be able to account for the increased 
He and N abundances in the progenitor stars.
The problems of convection and mixing, and of 
other modeling uncertainties (e.g., $n$-capture cross 
sections) requires further study in the context of 
$s$-process nucleosynthesis in intermediate-mass
AGB stars. 


%% Included in this acknowledgments section are examples of the
%% AASTeX hypertext markup commands. Use \url without the optional [HREF]
%% argument when you want to print the url directly in the text. Otherwise,
%% use either \url or \anchor, with the HREF as the first argument and the
%% text to be printed in the second.

\acknowledgments

We thank Rob Izzard, Dave Yong, and Khalil Farouqi for 
discussions about stellar nucleosynthesis, Matteo Cantiello
and Sabina Chita for discussions about stellar rotation, and
Robin Humble for help with the post-processing code. 
AIK acknowledges support from the Australian Research 
Council's Discovery Projects funding scheme (project number DP0664105);
partial support was provided by the Joint Theory Institute funded 
together by Argonne National Laboratory and the University of Chicago.
AIK also thanks the NWO and NOVA for money to visit Utrecht.
ML is supported by the NWO through the VENI fellowship scheme.
NCS is supported by an appointment to the NASA Postdoctoral Program 
at the Goddard Space Flight Center, administered by Oak Ridge 
Associated Universities through a contract with NASA.
HLD is supported by the National Science Foundation through
NSF grants AST 0406809 and 0708245.

\bibliography{apj-jour,/home/akarakas/biblio/library}
% using the macbook
%\bibliography{apj-jour,/Users/amanda/biblio/library}

%% Tables should be submitted one per page, so put a \clearpage before

\clearpage

\begin{table}[t]
\begin{center}
\caption{Details of the stellar models.\label{stellarmodels}}
\vspace{1mm}
\begin{tabular}{@{}cccccccccccc@{}} 
\tableline\tableline
 Mass &  $Z$  &  TPs & $T_{\rm He}^{\rm max}$ &  $T_{\rm bce}^{\rm max}$ 
   & M$_{\rm dred}$ & $M_{\rm env}$ & HBB? & C/O & \iso{12}C/\iso{13}C 
   & N/O  & He/H \\
\tableline
 3.0  &  0.02  &  26 & 302 & 6.75 & 8.1($-2$) & 0.676 & No  & 1.27 & 108  & 0.303 & 0.122 \\
 4.0  &  0.02  &  17 & 332 & 22.7 & 5.6($-2$) & 0.958 & No  & 0.99 & 76.5 & 0.336 & 0.118 \\
 5.0  &  0.02  &  24 & 352 & 64.5 & 5.0($-2$) & 1.500 & Yes & 0.77 & 7.84 & 0.542 & 0.133 \\  %%VW93
 5.0  &  0.02\tablenotemark{a} & 37 & 368 & 57.0 & 1.0($-1$) & 1.922 & Yes & 1.71 & 15.5 & 0.465 & 0.139 \\  %%R75
 6.0  &  0.02  &  37 & 369 & 83.1 & 5.8($-2$) & 1.791 & Yes & 0.38 & 10.8 & 1.251 & 0.150 \\
 6.5  &  0.02  &  40 & 368 & 86.5 & 4.7($-2$) & 1.507 & Yes & 0.40 & 11.6 & 1.205 & 0.154 \\ \tableline \tableline
 3.0  &  0.012 &  21 & 307 & 7.23 & 9.2($-2$) & 0.805 & No  & 2.80 & 190  & 0.366 & 0.117 \\
 6.5  &  0.012 &  51 & 369 & 90.0 & 6.5($-2$) & 1.389 & Yes & 0.76 & 10.4 & 2.503 & 0.146 \\ \tableline \tableline
 2.5  &  0.008 &  27 & 302 & 5.42 & 1.1($-1$) & 0.664 & No  & 4.17 & 392  & 0.307 & 0.109 \\ 
 5.0  &  0.008 &  57 & 366 & 80.8 & 1.7($-1$) & 1.795 & Yes & 0.93 & 7.48 & 5.937 & 0.128  \\
 6.0  &  0.008 &  68 & 374 & 89.6 & 1.2($-1$) & 1.197 & Yes & 1.40 & 8.80 & 5.799 & 0.135  \\ \tableline \tableline
 5.0  &  0.004 &  81 & 377 & 84.4 & 2.2($-1$) & 1.141 & Yes & 3.62 & 10.8 & 16.79 & 0.138  \\ \tableline
\tableline
\end{tabular}

\tablenotetext{a}{Computed with Reimers mass loss on the AGB. All other models have 
\citet{vw93} mass loss.}

\end{center}
\end{table}

\clearpage

\begin{table}[t]
\begin{center}
\caption{Surface abundance results for Zn through to Sr, taken at the
tip of the AGB from the last computed model. 
\label{abund}}
\vspace{1mm}
\begin{tabular}{crrrrrrrrrr}
\tableline\tableline
 Mass & PMZ & [Zn/Fe] & [Ge/Fe] & [Se/Fe] & [Br/Fe] & [Kr/Fe] & $\delta$\iso{86}Kr/\iso{82}Kr\tablenotemark{a} 
& [Sr/Fe] &  \multicolumn{1}{c}{$\Delta \,$Fe\tablenotemark{b}} \\
\tableline
\multicolumn{9}{c}{$Z = 0.02$, \citet{anders89} C, N and O} \\ \tableline
 3.0  & 0.0   & 0.001 & 0.004 & 0.004 & $< -$0.001 & 0.006 & $-$56.05 & 0.007 & $< 0.001$ \\
 3.0  & 0.001 & 0.071 & 0.263 & 0.346 & 0.299 & 0.477 & $-$693.3 & 0.700 & $-$0.00152  \\
 3.0  & 0.002 & 0.122 & 0.397 & 0.493 & 0.433 & 0.628 & $-$774.4 & 0.853 & $-$0.0021    \\
 4.0  & 0.0 &  0.004 & 0.011 & 0.009 & 0.003 & 0.010 & $-$25.82 & 0.007 & $-$0.0016 \\
 4.0  & 1($-4$) & 0.077 & 0.282 & 0.365 & 0.282 & 0.519 & $-$40.31 & 0.587 & $-$0.0026 \\
 5.0  & 0.0 & 0.012 & 0.031 & 0.026 & 0.011 & 0.028 & $-$10.85 & 0.011 & $-$0.0024 \\
 5.0  & 1($-4$) &  0.024 & 0.096 & 0.132 & 0.083 & 0.241 & 307.7 & 0.246 & $-$0.0024 \\
 5.0\tablenotemark{c} & 0.0 & 0.073 & 0.188 & 0.181 & 0.107 & 0.210 & 91.56 & 0.085 & $-$0.0090 \\
 6.0  & 0.0 & 0.038 & 0.106 & 0.104 & 0.051 & 0.131 & 92.09 & 0.055 & $-$0.0038 \\
 6.5  & 0.0 & 0.047 & 0.133 & 0.138 & 0.064 & 0.187 & 186.3 & 0.083 & $-$0.0037 \\ \tableline
\multicolumn{9}{c}{$Z = 0.012$, \citet{asplund05} C, N and O} \\ \tableline
 3.0 & 0.002   & 0.106 & 0.434 & 0.566 & 0.473 & 0.645 & $-$747.2 & 1.232  & $-$0.0028 \\
 6.5 & 0.0     & 0.042 & 0.113 & 0.111 & 0.054 & 0.136 & 55.60 & 0.053 & $-$0.0047 \\
 6.5 & 1($-4$) & 0.052 & 0.164 & 0.203 & 0.122 & 0.397 & 1171 & 0.300 & $-$0.0048 \\ \tableline
\multicolumn{9}{c}{$Z = 0.008$} \\ \tableline
 2.5 & 0.002 & 0.148 & 0.478 & 0.600 & 0.547 & 0.794 & $-$722.8 & 1.31 & $-$0.0040 \\
 5.0 & 0.0 & 0.163 & 0.380 & 0.401 & 0.302 & 0.506 & 355.3 & 0.264 & $-$0.0138 \\
 6.0 & 0.0 & 0.179 & 0.432 & 0.476 & 0.324 & 0.657 & 712.7 & 0.388 & $-$0.0098 \\ \tableline
\multicolumn{9}{c}{$Z = 0.004$} \\ \tableline
 5.0 & 0.0 & 0.403 & 0.813 & 0.897 & 0.734 & 1.176 & 1178  & 0.927 & $-$0.0216  \\
\tableline
\end{tabular}

\tablenotetext{a}{$\delta (^{i}X/^{\rm ref}X)$ notation is
defined according to $[ (^{i}X/^{\rm ref}X) / (^{i}X/^{\rm ref}X)_{\odot} - 1 ] 
\times 10^{3}$ where $^{\rm ref}X$ is a reference isotope.}
\tablenotetext{b}{Defined by $\log ( Y_{\rm final}/Y_{\rm initial})$, where Y is the
surface abundance in number fraction.}
\tablenotetext{c}{The model has Reimers mass loss on the AGB instead of \citet{vw93}.}

\end{center}
\end{table}

\clearpage

\begin{table}[t]
\begin{center}
\caption{Abundance results for O, Ne, P, S, Cl, and Ar.
\label{abund-2}}
\vspace{1mm}
\begin{tabular}{crrrrrrrrrr}
\tableline\tableline
 Mass &  $Z$ & $\epsilon$(O)\tablenotemark{a} & $\epsilon$(Ne) & Ne/O\tablenotemark{b} & $\epsilon$(P) & $\epsilon$(S) &
$\epsilon$(Cl) & $\epsilon$(Ar) & $\delta$\iso{38}Ar/\iso{36}Ar\tablenotemark{c} \\
\tableline
 initial & 0.02 & 8.935 & 8.103 & 0.181 & 5.591 & 7.277 & 5.143 & 6.570 & 0.000  \\ \tableline
 3.0  &  no PMZ   & 8.940 & 8.279 & 0.273 & 5.595 & 7.299 & 5.192 & 6.592 & 7.735 \\ % Delta O -0.023
 4.0  &  no PMZ   & 8.933 & 8.204 & 0.234 & 5.613 & 7.294 & 5.190 & 6.586 & 14.55 \\ %
 5.0  &  no PMZ   & 8.931 & 8.191 & 0.227 & 5.617 & 7.310 & 5.208 & 6.603 & 17.76 \\  %
 6.0  &  no PMZ   & 8.908 & 8.168 & 0.227 & 5.696 & 7.328 & 5.237 & 6.621 & 26.29  \\ % Delta 0 = -0.0859 
 6.5  &  no PMZ   & 8.905 & 8.166 & 0.228 & 5.635 & 7.330 & 5.212 & 6.623 & 25.18  \\ \tableline %Delta O =  -0.094
 initial  & 0.012 & 8.659 & 7.838 & 0.189 & 5.358 & 7.139 & 5.498 & 6.178 & 0.000  \\ \tableline
  3.0  &  no PMZ  & 8.656 & 8.086 & 0.336 & 5.393 & 7.165 & 5.530 & 6.205 & 17.00  \\ %Delta O = -0.0287
       &  0.002   & 8.666 & 8.666 & 0.486 & 5.522 & 7.165 & 5.534 & 6.204 & 41.34  \\ %Delta O =  -0.0187
  6.5  &  no PMZ  & 8.563 & 7.868 & 0.277 & 5.410 & 7.188 & 5.548 & 6.227 & 61.39  \\ \tableline %Delta O =  -0.153 
 initial  & 0.008 & 8.514 & 7.675 & 0.181 & 5.155 & 6.849 & 4.715 & 6.142 & 0.000  \\ \tableline
  2.5  & no PMZ   & 8.521 & 7.846 & 0.431 & 5.178 & 6.870 & 4.754 & 6.162 & 16.37  \\
       & 0.002    & 8.528 & 7.985 & 0.653 & 5.251 & 6.867 & 4.762 & 6.160 & 32.81 \\
  5.0  & no PMZ   & 8.419 & 7.817 & 0.313 & 5.454 & 6.899 & 4.886 & 6.192 & 92.67  \\  %Delta O =  -0.1503
  6.0  & no PMZ   & 8.317 & 7.781 & 0.364 & 5.471 & 6.908 & 4.861 & 6.201 & 63.44  \\ \tableline %Delta O =  -0.260
 initial  & 0.004 & 8.206 & 7.367 & 0.181 & 4.847 & 6.542 & 4.407 & 5.835 & 0.000  \\ \tableline
  5.0  & no PMZ   & 8.052 & 7.665 & 0.512 & 5.444 & 6.609 & 4.633 & 5.900 & 136.7  \\ \tableline % Delta O =  -0.228
\tableline

\tablenotetext{a}{$\epsilon$(Y) $= \log10 (Y/H) + 12$, where abundances are measured by number, 
and $H$ is the abundance of hydrogen.}
\tablenotetext{b}{The Ne/O ratio is calculated by mass, to allow comparison to the observations.}
\tablenotetext{c}{The $\delta$ notation was defined previously in Table~\ref{abund}.}

\end{tabular}
\end{center}
\end{table}

\clearpage


%% If you are not including electonic art with your submission, you may
%% mark up your captions using the \figcaption command. See the 
%% No more than seven \figcaption commands are allowed per page, 
%% so if you have more than seven captions, insert a \clearpage 

\clearpage

\clearpage

\clearpage

\clearpage

\end{document}